\documentclass[useAMS,usenatbib,usegraphicx]{mn2e}
\usepackage{subfigure}
\usepackage{float}
\usepackage{amssymb}
\usepackage{adjustbox}
\usepackage{soul}

\newcommand{\ms}{M$_{\odot}$}
\newcommand{\rs}{R$_{\odot}$}

\newcommand{\virgopen}{``}
\newcommand{\virgclose}{''}

\renewcommand*{\thesubfigure}{} 

\title{The effect of a wider initial separation on common envelope binary interaction simulations}
\author[Iaconi et al.]{Roberto Iaconi $^{1}$\thanks{E-mail:roberto.iaconi@students.mq.edu.au}, Thomas Reichardt $^{1}$, Jan Staff $^{1,2}$, Orsola De Marco $^{1}$, \newauthor Jean-Claude Passy $^{3}$, Daniel Price$^{4}$, James Wurster$^{4,5}$ and Falk Herwig$^6$\\
$^{1}$Department of Physics \& Astronomy, Macquarie University, Sydney, NSW 2109, Australia\\
$^{2}$Department of Astronomy, the University of Florida, Gainesville, FL 32611, USA\\
$^{3}$Argelander Institute f\"{u}r Astronomie, Bonn Universit\"{a}t, Bonn, Germany\\
$^{4}$Monash Centre for Astrophysics and School of Physics and Astronomy, Monash University, Clayton, VIC 3800, Australia\\
$^{5}$Astrophysics Group, School of Physics, University of Exeter, Stocker Road, Exeter EX4 4QL\\
$^{6}$Department of Physics and Astronomy, University of Victoria, Victoria BC, Canada}

\begin{document}

\date{Submitted to MNRAS}

\pagerange{\pageref{firstpage}--\pageref{lastpage}} \pubyear{2002}

\maketitle

\label{firstpage}

\begin{abstract}
We present hydrodynamic simulations of the common envelope binary interaction between a giant star and a compact companion carried out with the adaptive mesh refinement code {\sc enzo} and the smooth particle hydrodynamics code {\sc phantom}. These simulations mimic the parameters of one of the simulations by Passy et al., but assess the impact of a larger, more realistic initial orbital separation on the simulation outcome. We conclude that for both codes the post-common envelope separation is somewhat larger and the amount of unbound mass slightly greater when the initial separation is wide enough that the giant does not yet overflow or just overflows its Roche lobe. {\sc phantom} has been adapted to the common envelope problem here for the first time and a full comparison with {\sc enzo} is presented, including an investigation of convergence as well as energy and angular momentum conservation. We also set our simulations in the context of past simulations. This comparison reveals that it is the expansion of the giant before rapid in-spiral and not spinning up of the star that causes a larger final separation. We also suggest that the large range in unbound mass for different simulations is difficult to explain and may have something to do with simulations that are not fully converged.  
\end{abstract}


\section{Introduction}
\label{sec:introduction}

The common envelope (CE) interaction is a short phase of the interaction between two stars in a binary system characterised by the dense cores of the two objects orbiting inside their merged envelopes. It was first described by \citet{Paczynski1976}, but see also \citet{Ivanova2013} and references therein. During this phase, orbital energy and angular momentum are transferred to the gas of the envelope, which can become unbound from the potential well of the system, leaving behind a close binary. In cases when the envelope is not unbound, a merger results. 

The CE phase is thought to be the main evolutionary channel that leads to all the evolved compact binaries. Post-CE compact binaries can eventually merge on longer time-scales. In addition to compact binaries, mergers inside the CE can take place. In this case energy and angular momentum from the orbit are entirely dissipated into the envelope, which may be not entirely ejected. Objects and phenomena resulting from these scenarios include type Ia supernovae, low and high mass X-ray binaries, double neutron star and double black holes. A full physical description of all binary interaction scenarios, including the CE phase, is essential in constructing state-of-the-art stellar population synthesis models to understand the rates at which compact binaries and mergers form (see \citealt{Toonen2014} and references therein for an exhaustive review of both binary evolution scenarios including CE and their rates).  Hydrodynamics simulations are an essential tool to investigate the physics of the CE phase and determine the outcome of CE interactions as a function of initial binary parameters.

Past efforts have tried to reproduce numerically CE interactions with different codes (e.g., \citealt{Rasio1996}, \citealt{Sandquist1998}, \citealt{Sandquist2000}, \citealt{Passy2012}, \citealt{Ricker2012}, \citealt{Nandez2014}), but failed to reproduce the post-CE systems observed. Primarily, simulations fail to unbind the entire envelope. While the envelope is lifted away from the in-spiralling binary, the majority of it is not unbound (e.g., \citealt{Passy2012}). Recently \citet{Ivanova2015} and \citet{Nandez2015} reported that adding recombination energy in their simulations achieves the unbinding of the envelope under at least a certain combination of parameters. 

Current simulations are limited in one way or another. The range of physical phenomena taken into consideration is still very limited (e.g., the effect of magnetic fields is possibly important; \citealt{Regos1995}, \citealt{Nordhaus2007}, \citealt{Tocknell2014}, \citealt{Ohlmann2016b}). In addition, the initial conditions of the simulations are often non-physical. For example, many simulations start with the companion on or close to the surface of the primary (\citealt{Passy2012} and \citealt{Ricker2012}). Despite the growing number of simulations, binary parameter space is still sparsely covered. Additionally, different numerical techniques are used, e.g., unigrid (\citealt{Passy2012}), adaptive mesh refinement (AMR; \citealt{Ricker2012}), smoothed particle hydrodynamics (SPH; \citealt{Nandez2014}) and unstructured mesh (\citealt{Ohlmann2016}), but only seldom benchmark comparisons exist (e.g., \citealt{Passy2012}). Finally, the resolution of the simulations is always relatively low (but see the improvement in the latest simulations by \citealt{Ohlmann2016}) and the available convergence tests are never exhaustive enough, due to the substantial computational demands of these simulations, to convince one that resolution does not play a part in the outcome of the simulations. Thus, determining the effect that individual aspects of the simulations have on the results is a way to provide insight into which of the effects has the largest impact on the simulation's outcome. 

Here we analyse the effect of the initial orbital separation on the final outcome of CE simulations by carrying out a set of simulations that parallel one of the simulations carried out by \citet[][hereafter P12]{Passy2012}, where a 0.88~\ms\ red giant branch (RGB) star interacts with its 0.6~\ms\ compact companion. In their simulation the companion was initially placed near the surface of the giant. In one of our simulations we place instead the companion at the approximate largest distance from which an orbiting companion is likely to be brought into Roche lobe contact with a giant. It is expected that prior to the start of the CE in-spiral phase, tidal forces will redistribute orbital energy and angular momentum from the orbit to the primary. Eventually the primary would overflow its Roche lobe and start mass transfer to the companion, finally resulting in the fast CE in-spiral. These phases are expected to induce envelope rotation and expansion, changing the overall distribution of the envelope and lowering its binding energy. The envelope would be lighter and easier to unbind, but the overall strength of the gravitational drag (\citealt{Ostriker1999}) may be smaller because of relatively lower densities and smaller velocity contrasts. It is therefore not clear {\it a priori} what effect a larger initial separation would have on the simulation. 

The effect of a rotating giant on the CE outcome could only be gauged by \citet{Sandquist1998} who carried out side-by-side simulations with rotating and non-rotating giants and determined that the outcome does not vary much. \citet{Rasio1996}, \citet{Ricker2012} and \citet{Ohlmann2016} all used a rotating giant, but did not compare their results with a non-rotating case. In addition, while \citet{Rasio1996} stabilised the rotating giant, none of the other studies did, introducing doubts as to the impact of the giant rotational expansion on the results. Finally, all simulations started at a separation such that the giant was already overflowing its Roche lobe and thus could not gauge the effects of a more gradual expansion of the giant envelope.

In line with the work of P12 we carry out our simulations with grid (in AMR mode) and SPH codes. In so doing we continue to compare different numerical techniques while making the most of what each has to offer. The SPH code we use, {\sc phantom} \citep{Price2010, Lodato2010}, has never been used for CE interaction simulations before, hence this work serves also to introduce {\sc phantom} to this problem.

This paper is organised as follows. In Section~\ref{sec:setup} we explain the simulations' setup. In Section~\ref{sec:results} we analyse the outputs of our simulations, focusing on the evolution of the orbital separation in Section~\ref{ssec:separation}, the distribution of the envelope in Sections~\ref{ssec:gas} and \ref{ssec:bulges}, the gravitational drag during the interaction in Section~\ref{ssec:density} and the energetics and the numerical behaviour of the codes in Section \ref{ssec:conservation}. In Section~\ref{sec:otherwork} we set our results in the context of all previous simulations while in Section~\ref{sec:conclusion} we conclude.

\section{Simulations setup}
\label{sec:setup}

We use two different codes to simulate our physical problem: {\sc Enzo} \citep{OShea2004,Bryan2014}, an Eulerian code with  AMR and {\sc phantom} (\citealt{Price2010}, \citealt{Lodato2010}) an SPH code.

{\sc Enzo} is a parallel 3D hydrodynamic code including self gravity, originally developed for cosmological simulations, which has been adapted for CE simulations as described in P12. {\sc enzo} had already AMR capabilities when P12 performed their simulation, but they were not available for CE simulation. Therefore, they performed their simulations with a static, uniform grid. However, given the most recent updates applied to {\sc Enzo} (\citealt{Passy2014}), we used the AMR capabilities of the code, which guarantee better resolution where needed and a better usage of computational resources.

Our simulation has been run with a cubic box of $863$~\rs~$=4$~AU on a side and a coarse grid resolution of $128$ cells per side. We adopt two levels of refinement with a refinement factor of two (i.e., when a cell is refined it is divided by two along each dimension),in this way the smaller cell size is $1.68$~\rs, as was the case in the $256^3$ simulations of P12. The refinement criterion is based on cell gas density. Cell densities above $1.38 \times 10^{-8}$~g~cm$^{-3}$ dictate a cell division. Additionally {\sc Enzo} adaptively de-refines the zones where a cell and its surrounding region no longer satisfy the refinement criterion. For our choice of the smoothing length (see below), two levels of refinement are the minimum to obtain a stable giant model with the best possible energy conservation.

\citet{Ricker2012}, who carried out the only other CE simulations adopting an Eulerian AMR approach, use a computational domain of $575$~\rs~$=2.7$~AU with $9$ levels of refinement, obtaining a smallest cell length of $0.29$~\rs, approximately $6$ times smaller than the value we obtain in this work.

As we will explain in Section~\ref{ssec:single} and Section~\ref{ssec:binary}, we use point-masses, interacting only gravitationally with both gas and other particles, to model the primary core and the companion. This point-mass potential is  smoothed according to the prescription of \citet{Ruffert1993}. To ensure  reasonable energy conservation, our smoothing length is equal to $3$ times the smallest cell size, as this was found to be the optimal value by \citet{Staff2016a}, who monitored the energy conservation in their CE {\sc Enzo} simulations as a function of smoothing length. A smoothing length of $1.5$ times the smallest cell size, as used by \citet{Sandquist1998} and P12, results in a non-negligible energy non-conservation in our simulations for this particular case. Increasing the smoothing length reduces the strength of the gravity over a larger volume around the point masses. Our choice for the smoothing length ($3 \times$[smallest cell size] $\simeq 5$~\rs) yields a radius inside which gravity is smoothed that is double that of P12's $256^3$ simulations.

{\sc phantom} is a shared memory (OpenMP) parallelised, 3D SPH code. {\sc phantom} was originally designed to model star formation, but it has been expanded to simulate different types of astrophysical problems thanks to its modularity. We modified {\sc phantom}, allowing the code to setup 3D stellar models based on 1D stellar evolution codes radial profiles and then to create binary systems for CE simulations. 

Our {\sc phantom} simulations have been run with variable total numbers of particles to test for convergence (see Section~\ref{ssec:phantom_convergence_test}). However, the main simulations we use for our results have been carried out with $1$ million particles. Similarly to the {\sc Enzo} procedure, we use point-masses (called sink particles in the {\sc phantom} nomenclature), interacting only gravitationally with both gas and SPH particles, to model the primary core and the companion (see Secs.~\ref{ssec:single} and \ref{ssec:binary}). Both core and companion particles were given a softening length\footnote{The softening length in {\sc phantom} is equivalent to the smoothing length in {\sc Enzo}. {\sc phantom} reserves the term ``smoothing length" for the size of the smoothing kernel, such that each SPH particle has a smoothing length.} of 3~\rs, irrespective of the total number of particles used. 

The methodology followed to simulate our CE interaction consists of two main phases and is described in the following sections.

\subsection{Single star setup and stabilisation}
\label{ssec:single}

As in P12 we model our binary system as an RGB primary and a smaller companion with comparable mass, identifiable with a main sequence star or a compact object such as a white dwarf. The resolution is not sufficient to resolve the primary's core, nor the companion, so we model them as dimensionless point-masses. The companion mass is $M_2=0.6$~\ms \ (this choice will be discussed in Section 2.2). The primary star is an extended object whose envelope is well resolved. We use the same initial model as in P12: a star with an initial mass of $1$~\ms~evolved to the RGB with the 1D stellar evolution code {\sc Evol} (\citealt{Herwig2000}). At this stage of the evolution the star has a radius of $R_1 = 83$~\rs, a total mass of $M_1 = 0.88$~\ms~and a core mass of $M_c = 0.392$~\ms.

The relevant {\sc Enzo} physical quantities are interpolated from the 1D model to the 3D domain. Due to the limited resolution of the 3D code the interpolation process results in a mass deficit that coincides almost exactly with the mass of the core. The addition of a point mass representing the missing mass therefore completes the stellar structure. Moreover, because of the limited resolution, the surface of the star is poorly matched to the steep gradients typical of stellar atmospheres, therefore the part of the simulation box not occupied by the star is filled by low density medium with a density equal to $10^{-4}$ times the density of the surface layer of the primary. To match the pressure of the stellar atmosphere this medium has a high temperature ($\simeq 10^8$~K). However, the stellar model is not in perfect hydrostatic equilibrium in the grid due to the higher resolution adopted by {\sc Evol} and its more realistic equation of state that takes micro physics into account, while {\sc Enzo} has an ideal gas equation of state with $\gamma = \frac{5}{3}$. The primary therefore has to be stabilised by damping at each time step the velocities that develop in the grid. This stabilisation is carried out for $10$ dynamical times. The stability of the model in the new grid is then verified by letting the simulation run without damping for $10$ additional dynamical times, where our RGB star dynamical time is $21$~days.

\begin{figure}
\centering     
\subfigure[]{\label{fig:b}\includegraphics[scale=0.45, trim=0.8cm 0.0cm 0.0cm 0.0cm]{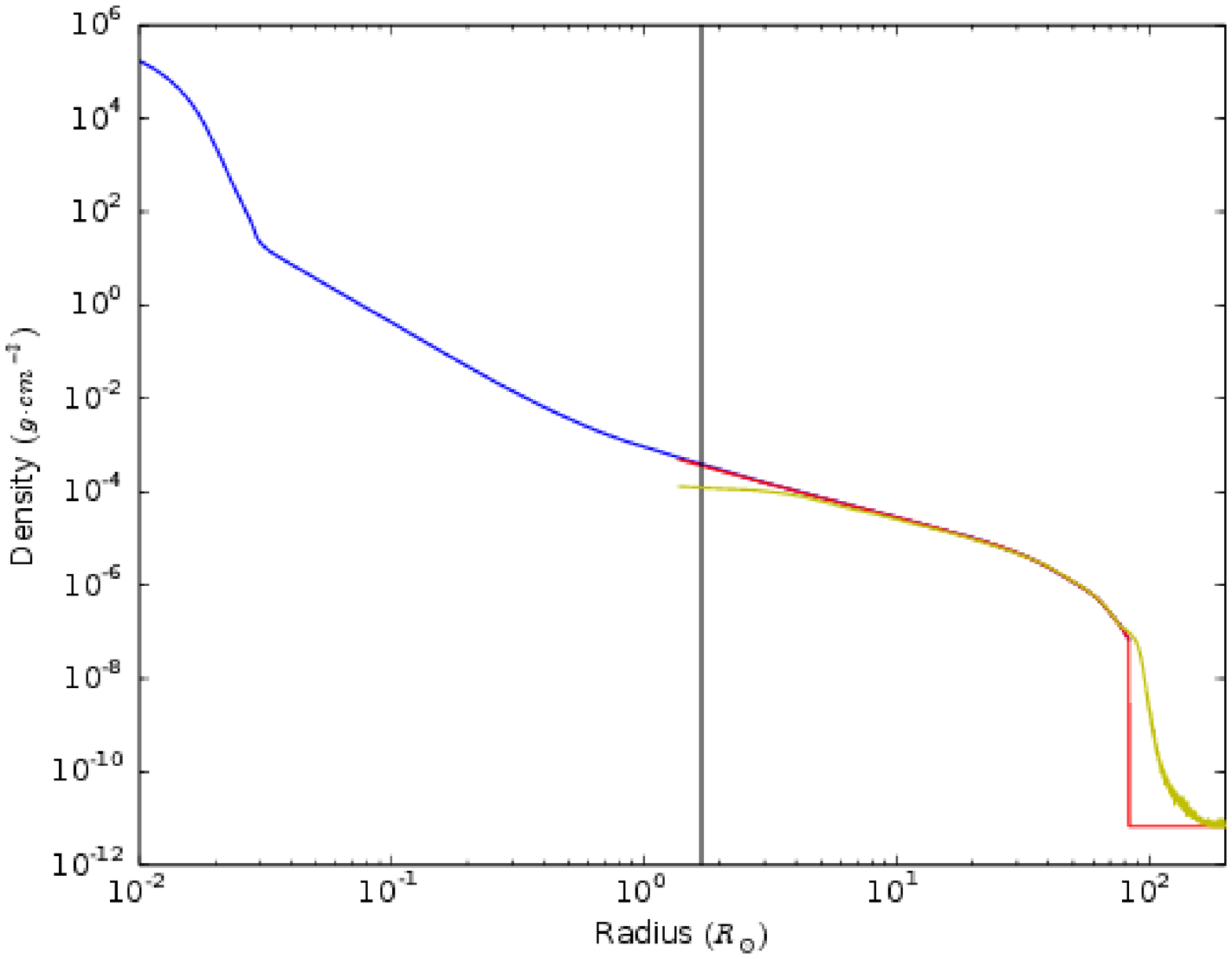}}
\subfigure[]{\label{fig:a}\includegraphics[scale=0.45, trim=0.8cm 0.0cm 0.0cm 0.0cm]{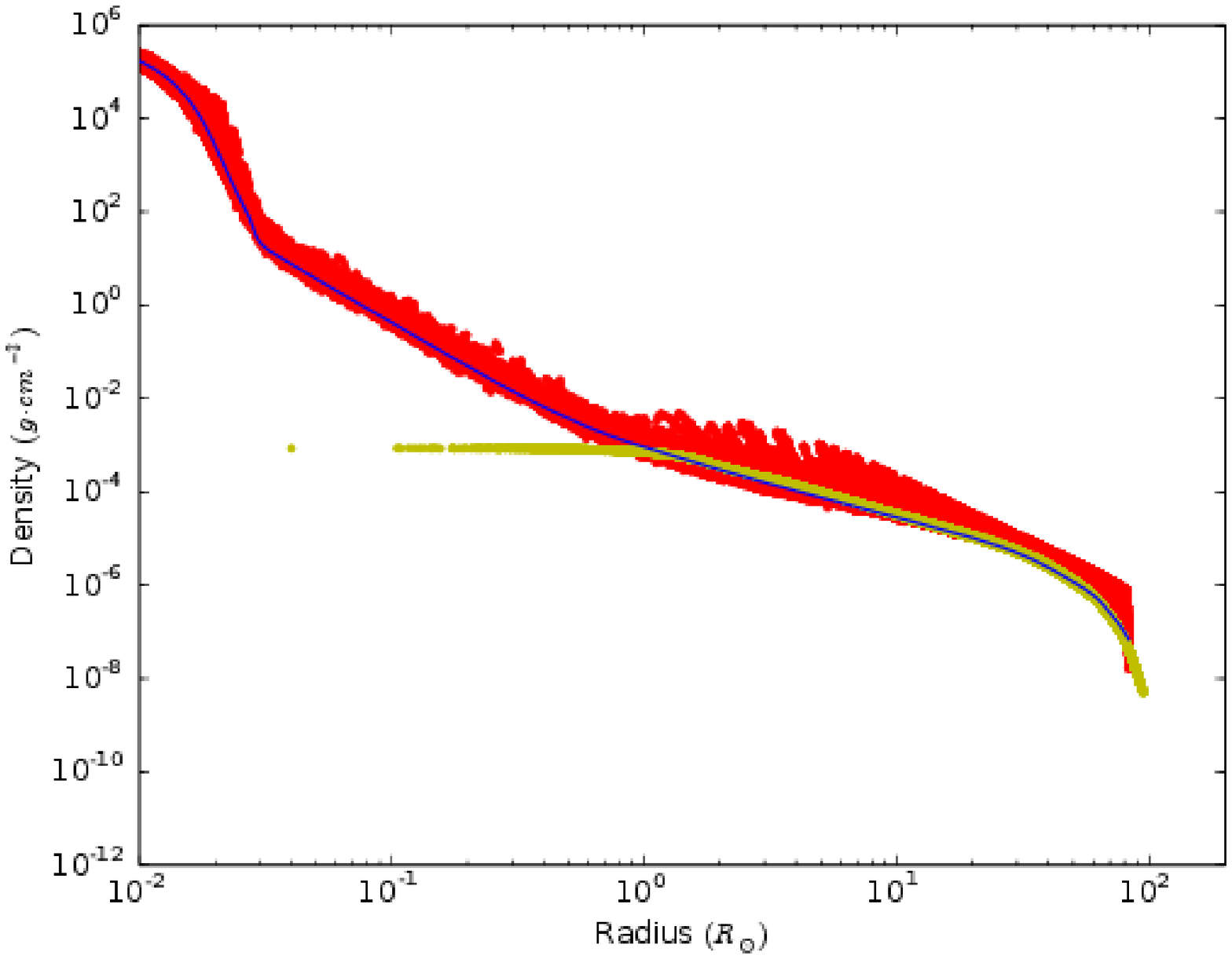}}
\caption{\protect\footnotesize{Upper panel: radial density profiles of the primary RGB star used in our {\sc Enzo} simulation, calculated with the 1D {\sc Evol} code (blue), after mapping it in the {\sc Enzo} computational domain but before the stabilisation process (red) and after stabilisation (yellow). The change in slope at a radius of $3\times 10^{-2}$~\rs \ marks the core-envelope boundary of the 1D model, while the vertical line shows the size of an {\sc Enzo} cell at the deepest level of refinement. Lower panel: same as the top panel, but for the {\sc phantom} code using 2.3 millions particles. Note that while for {\sc Enzo} we perform a radial average, showing a single density value at each radius, for {\sc phantom} we plot the density of all the SPH particles at a given radius. As a result the red curve is not simply a line. The lowest density in the {\sc phantom} simulation is larger than for {\sc enzo} because of the lack of a low-density ``vacuum".}}
\label{density_profile_1D_vs_3D}
\end{figure}

At the end of this process the initial 3D stellar model is relaxed with respect to the 1D model as showed in Figure \ref{density_profile_1D_vs_3D} (upper panel). The sharp density jump at the edge of the star has been smoothed by the stabilisation process, and the star is now slightly larger. The contour of density of $10^{-11}$~g~cm$^{-3}$ has a radius $100$~\rs. The central density is also sightly reduced, but overall the original structure of the star is mostly preserved.

To verify the stability of the model more quantitatively than previously done (\citealt{Sandquist1998}, \citealt{Ricker2012}, P12), the velocities that develop have been compared to global and local velocity scales, such as the local sound speed and the dynamical velocity, $v_{dyn,1} = {R_1}/{t_{dyn,1}} \simeq R_1 (G \langle \rho_1 \rangle)^{\frac{1}{2}}$, where $t_{dyn,1}$ is the dynamical time of the primary, $G$ is the gravitational constant and $\langle \rho_1 \rangle$ is the average density of the star. Additionally, we also compare the gas velocities in the frame of reference of the primary to the orbital velocities of the binary system in the frame of reference of the center of mass (see Section 2.2). At each step during the relaxation at most 7 per cent of the cells had velocities exceeding the lowest of the velocity limits discussed above. Hence we expect the contamination of the CE interaction by the spurious motions of the primary envelope to be negligible.

In {\sc phantom} we map the same 1D stellar model, but in this case the SPH particles are distributed so as to reproduce the entire stellar mass distribution, inclusive of the core. This generates a very high particle density at the location of the core that would slow down the simulation excessively. We therefore  approximate the core of the giant using a sink particle set up to accrete all SPH particles within a radius of $0.03$~\rs. This quickly generated a ``core" with the correct mass ($M_c$=$0.392$~\ms). Note that the number of particles mentioned for all the {\sc phantom} simulations in the following sections is the actual number of particles after the accretion process (e.g., the convergence test using $2.3 \times 10^6$ particles described in Section~\ref{ssec:phantom_convergence_test} was actually initialised with $\simeq 4 \times 10^6$ particles).
The giant was then damped and stabilised as was done for {\sc enzo}. The profiles of the star as mapped in {\sc phantom} is shown in Figure~\ref{density_profile_1D_vs_3D} (lower panel).

\renewcommand*{\thesubfigure}{} 

\subsection{Binary system setup}
\label{ssec:binary}
 
The companion has a mass $M_2 = 0.6$~\ms \ for both the {\sc enzo} and the {\sc phantom} simulations, selected among those simulated by P12, also based on the fact that their $0.6$~\ms \ companion simulations were converged for the coarse grid resolutions we are using. However, the {\sc Enzo} and {\sc phantom} simulations differ for the initial separation of the binary.

In {\sc Enzo} the orbital separation was the largest that would result in the evolution of the orbital elements and eventually in a CE: $a=300$~\rs\ (corresponding to a period of $496$~days~$=1.36$~yr). This value also corresponds to the approximate maximum orbital separation from which a tidal capture of the companion may take place within the evolution of a star similar to our primary: \citet{Madappatt2016} shows that a 1.5~\ms \ star grows to have a maximum RGB radius of 130~\rs \ and can engulf a 0.15~\ms \ companion that orbits as far as 2.5 times that radius. Hence it is reasonable that our star with a radius of $\sim$100~\rs, can succeed in capturing tidally a companion that is as far as approximately 300~\rs. The system was placed in circular orbit, where we gave the RGB star a Keplerian velocity $v_{1} \simeq 12.4$~km~s$^{-1}$ and the companion point particle a velocity $v_{2} \simeq 18.2$~km~s$^{-1}$, with the point mass core of the primary coinciding with the centre of the box.

In our simulation the primary is driven into Roche lobe contact (the Roche lobe radius of the primary is 124~\rs\ at an orbital separation of $300$~\rs, using the approximation of \citealt{Eggleton1983}, but noting it to be valid in the case of synchronised orbits, which is not our case) and eventually a CE interaction by the pre-contact tidal interactions in a relatively short time-scale ($\simeq 1.5$~yr, see Section~\ref{ssec:separation}), much shorter than realistic tidal interaction time-scales. Tidal interaction simulations performed by \citet{Madappatt2016} show in fact that the time-scale for the engulfment of a $0.15$~\ms \ companion by a $1.5$~\ms \ primary initially orbiting at $\simeq 300$~\rs \ are of the order of 100\,000~yr. Although this system is slightly different from the one simulated here, a tidal interaction time not too dissimilar is expected in our case. The reason for this difference is that the strength of the interaction is sensitive to departures of the stellar envelope distribution from spherical. Inserting the companion in the computational domain generates a small distortion of the primary's envelope resulting in a set of oscillations, which exert a relatively strong tidal force. Paradoxically, this larger than average tide results in shortening of the orbital separation within reasonable computational times, something that would not be so if the tide were better reproduced. For more discussion on this topic see Sec.~\ref{ssec:bulges}.

We do not apply any initial rotation to the primary. However, we achieve a spinning star by spin-orbit interaction. This means that the total angular momentum in the system, which is increasingly transferred from the orbit to the envelope of the giant, is approximately that which would be expected for this system (see Sec.~\ref{ssec:conservation}).  

We carried out two main {\sc phantom} simulations. The first one has similar parameters to that carried out by P12 with a companion mass of $0.6$~\ms \ and is used as a verification step to ensure that {\sc phantom} performs similarly to the other codes we have used. This simulation's outcomes were compared directly with the SPH simulation ``SPH2", which in that study was carried out  with the SPH code {\sc snsph} (\citealt{Fryer2006}) using 500\,000 SPH particles. Additionally, for this binary configuration, we carried out a resolution test, described in Section~\ref{ssec:phantom_convergence_test}.

The second {\sc phantom} simulation has a larger initial separation, to corroborate whether a larger initial separation promotes a wider final separation. The initial separation we use in this case is $218$~\rs, the distance at which the primary fills its Roche lobe. Ideally we would have used a larger separation of $300$~\rs, like for the {\sc Enzo} simulation discussed above. However, the orbital evolution of a {\sc phantom} simulation with an initial separation of $300$~\rs \ was too slow to reach the common envelope phase in reasonable computational times. This is due to the  stability of SPH simulations to surface deformations (\citealt{Springel2010}; see also our discussion in Section~\ref{ssec:bulges} and Figure~\ref{tidal_bulge_mass}).

\citet{Rasio1996}, \citet{Nandez2014} and \citet{Nandez2015} stabilise their giants in the co-rotating frame of the binary, while slowly decreasing the orbital separation to the desired value (for a more detailed description see \citealt{Rasio1996}). We do not apply this additional stabilisation in our simulations. Similarly to us, the simulations of \citet{Sandquist1998}, \citet{Sandquist2000} and \citet{Ricker2012}, all starting with a separation larger than the radius of the primary (see Table~\ref{comparison_with_previous_works_data}), did not stabilise their giants in the co-rotating frame.

\subsection{{\sc phantom} convergence test}
\label{ssec:phantom_convergence_test}
\begin{figure} 
\centering
\includegraphics[scale=0.45, trim=0.8cm 0.0cm 0.0cm 0.0cm]{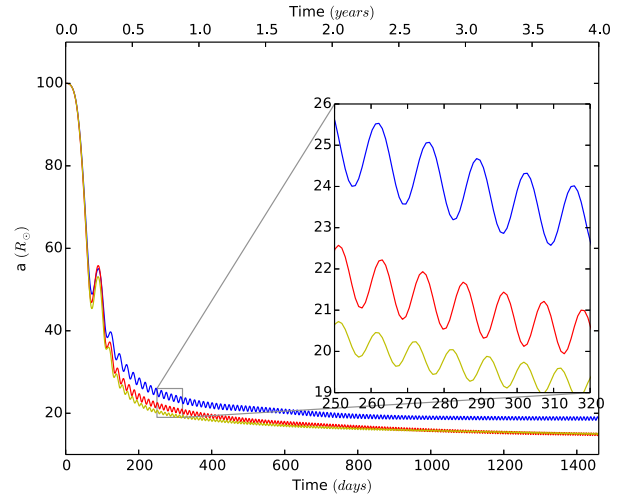}
\caption{\protect\footnotesize{Evolution of the separation, $a$, between the two particles representing the core of the primary and the companion, used to show the convergence for the {\sc phantom} code. The simulation reproduces the one from P12 with the same companion's mass as this work ($M_2 = 0.6$~\ms). The number of SPH particles used is: $2.3 \times 10^4$ (blue), $2.3 \times 10^5$ (red), $2.3 \times 10^6$ (yellow). The inset shows a $10 \times$ zoom on the end of the rapid in-spiral phase.}}
\label{sph_convergence_test}
\end{figure}

Since {\sc phantom} was used here for the first time to simulate a CE interaction, we carried out a convergence test to better understand the behaviour of the code at different resolutions. We used $3$ resolutions: $23,000$, $230,000$, and $2.3$ million particles, and we show the evolution of the orbital separation in the three cases in Figure~\ref{sph_convergence_test}. The factor of $10$ difference between the resolutions is just larger than the minimum resolution step needed for such a test. While this test shows that we have not yet achieved formal convergence, the change in orbital evolution with resolution is much smaller between the higher two resolutions than between the lower two, indicating converging behaviour.

\section{Results}
\label{sec:results}

\subsection{Orbital Separation}
\label{ssec:separation}

\begin{figure}
\centering     
\subfigure[]{\label{fig:b}\includegraphics[scale=0.4]{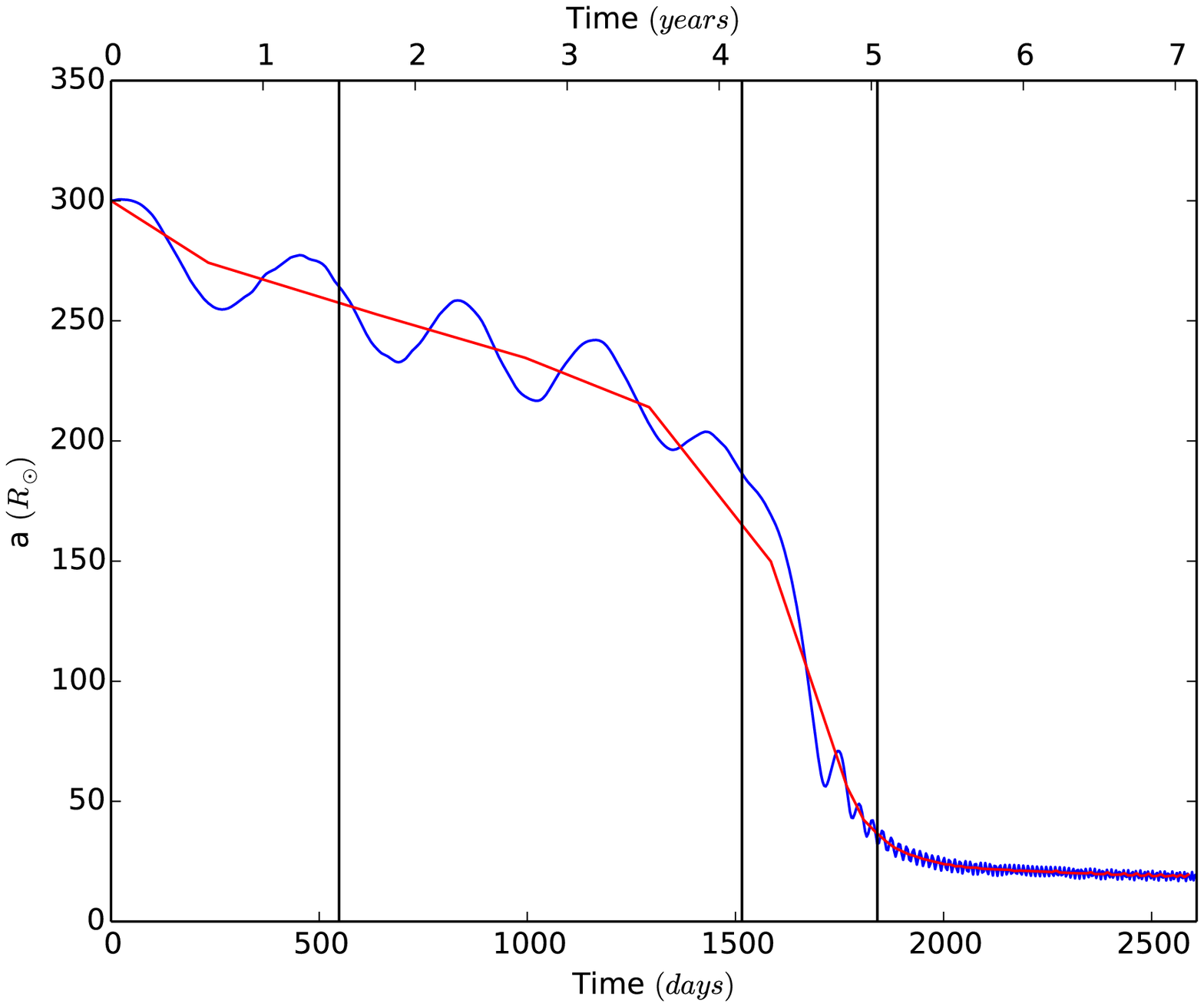}}
\subfigure[]{\label{fig:a}\includegraphics[scale=0.4]{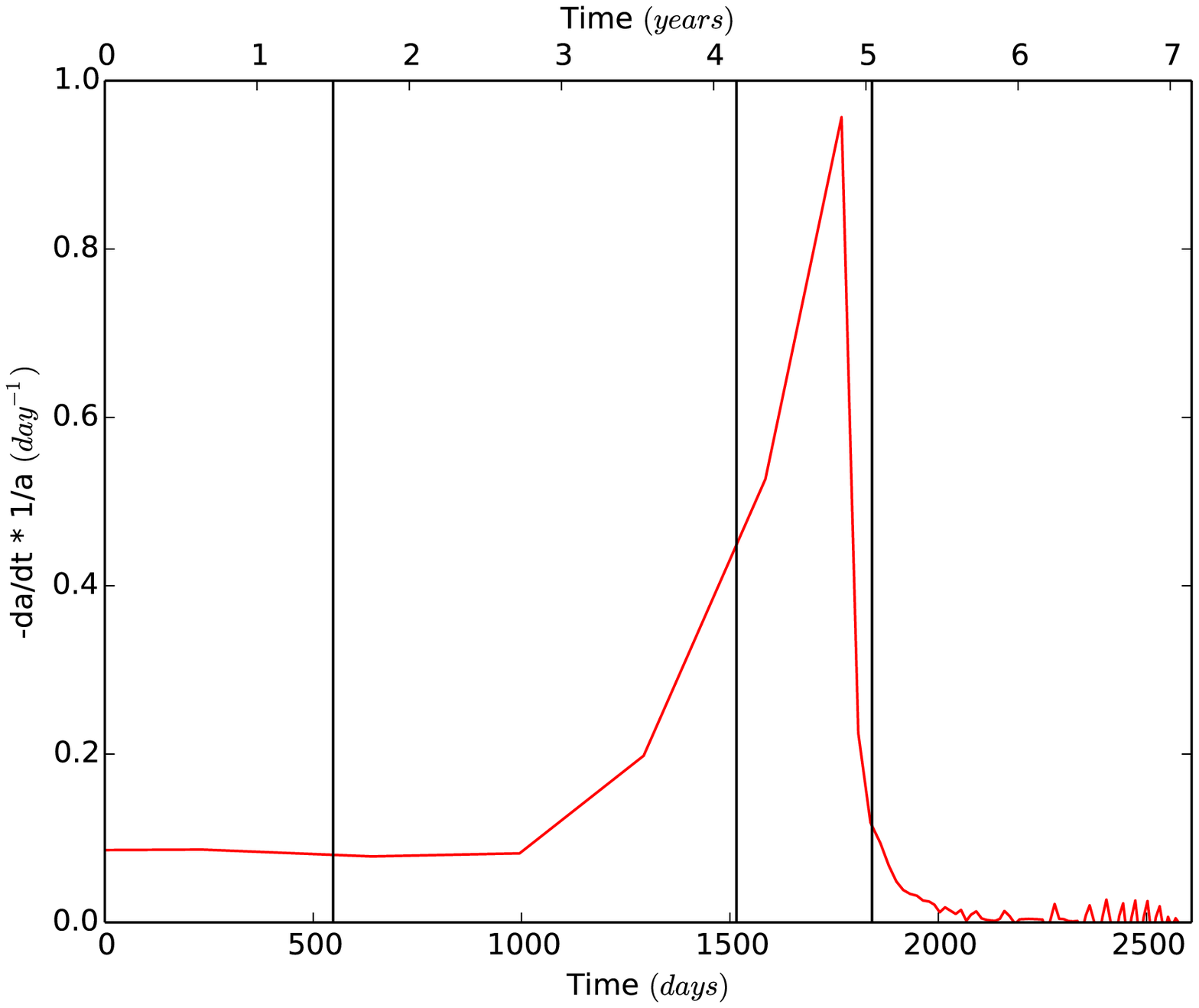}}
\caption{\protect\footnotesize{Upper panel: evolution of the separation, $a$, between the two particles representing the core of the primary and the companion, over the whole simulation time for the {\sc Enzo} simulation. The blue line represents the actual separation computed every $0.01$~year. The red line represents the separation averaged over one orbital cycle. The black vertical lines represent, from left to right, the beginning of mass transfer, the beginning of the fast in-spiral phase and the end of the fast in-spiral phase. Lower panel: evolution of the orbital decay, computed on the separation averaged over one orbital cycle for the same simulation.}}
\label{separation_vs_time}
\end{figure}

For our {\sc Enzo} simulation the separation between the point masses as a function of time and the orbital decay rate are shown in Figure~\ref{separation_vs_time}.  To determine the time when the mass transfer phase begins, we calculated the Roche lobe surface around the primary using the total potential field computed in the simulation. Then, we checked whether the cells contained within the primary's Roche lobe, including the first cell near the inner Lagrangian point in the companion's Roche lobe, have a density greater than the vacuum's density ($6.93 \times 10^{-12}$~g~cm$^{-3}$). Computed in this way, the beginning of the contact phase takes place after about $547$~days $=1.5$~yr from the beginning of the simulation. During this pre-contact phase, the orbital separation has been reduced from $300$ to $265$~\rs, at which point the primary's Roche lobe radius is $108$~\rs, similar to the stellar radius at the start of the simulation.

\renewcommand*{\thesubfigure}{} 

The mass transfer phase lasts until the companion is engulfed in the envelope of the primary, at which point the rapid in-spiral phase begins. We define the start of the rapid in-spiral phase as the time when the equipotential surface passing through the outer Lagrangian point L$_2$ has a density greater than the vacuum's density in each of its cells. This condition is satisfied after about $1515$~days, or $4.2$~yr, from the beginning of the simulation.

The rapid in-spiral phase is observed as a steepening of the the separation vs time curve which denotes a regime change. This phase lasts $324$~days and ends $1840$~days ($5.0$~yr), from the beginning of the simulation, when the orbital separation stabilises. We have used the same criterion as P12 and \citet{Sandquist1998}, who defined the end of the rapid in-spiral phase when $-\dot{a} < 0.1(-\dot{a}_{max})$, where $\dot{a} = da/dt$. This point is somewhat arbitrary because it depends on how steep the in-spiral is and, in our simulations, the in-spiral is much steeper than that witnessed in the simulations of \citet{Sandquist1998} and P12. This can be seen by comparing our Figure~\ref{separation_vs_time}, lower panel with their figures 4 and 5, respectively. As a result, the separation vs. time curve is slightly steeper than the ones of \citet{Sandquist1998} and P12 at the point when we define the end of the rapid in-spiral using this criterion.

The rapid in-spiral phase in our simulation lasts approximately 10 per cent longer than for the equivalent simulation of P12, and longer still if we acknowledge that at the end of the in-spiral phase as defined above, the separation is still reducing considerably. This could be due to the fact that our donor star is puffed up by the interactions in the previous phases, hence it is less dense. The delayed rapid in-spiral and its longer duration are in line with the results obtained by P12 in their simulations with the companion star slightly away from the primary surface rather than in contact. 

The orbit starts to become elliptical during the rapid in-spiral phase. Using the maxima and minima in the orbital separation evolution after the end of the rapid in-spiral phase, we obtain an eccentricity $e = 0.12$, in agreement with what was obtained by P12.

The final separation achieved ($a_f$) is a crucial output of the CE simulations. P12 identified that CE simulations have final separations that not only tend to be larger than observed (\citealt{Zorotovic2010} and \citealt{DeMarco2011}), but that depend on the companion/primary mass ratio ($q$), a tendency not seen in the observations. By using the average separation (red line in Figure \ref{separation_vs_time}) we estimated the value of the separation reached at the end of the rapid in-spiral phase in our {\sc Enzo} simulation to be $36$~\rs, using the criterion described above and $20$~\rs\ if we take the average value at the end of the simulation (see Table~\ref{comparison_with_previous_works_data}, where we report the initial conditions and final outcomes for all past CE simulations including at least a giant).  The separation at the end of the simulation is $\simeq 4$ times the smoothing length, indicating that the end of the in-spiral is not affected by the smoothing-length and resolution. 
Our values of the final separation are larger than those of P12, which were $19$ and $16$~\rs, for the criterion-defined and final separations, respectively.  In other words, the final separation is larger by 25 per cent for the {\sc enzo} simulation starting with a larger initial separation. 

\begin{figure} 
\centering
\includegraphics[scale=0.45, trim=0.8cm 0.0cm 0.0cm 0.0cm]{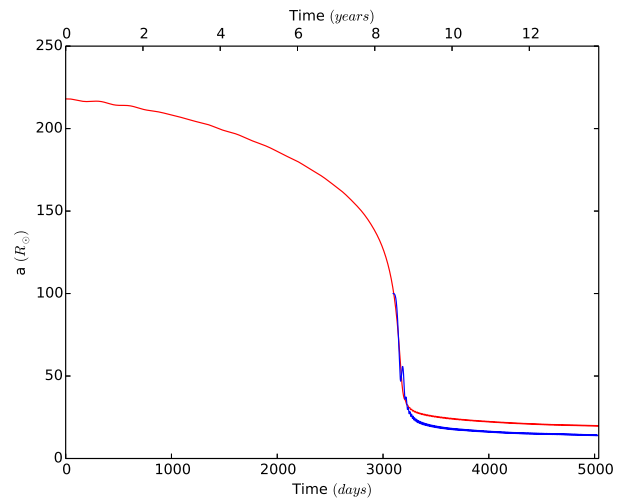}
\caption{\protect\footnotesize{Evolution of the separation, $a$, between the the two particles representing the core of the primary and the companion for the {\sc phantom} simulations with initial separations of $100$~\rs \ (blue curve) and $218$~\rs \ (red curve). For a clearer comparison of the final separation the blue line has been shifted forward in time by $3096$ days, which is the time when the orbital separation of that simulation reaches $100$~\rs.}}
\label{sph_separations}
\end{figure}

For our first {\sc phantom} simulation, carried out with the same initial configuration as the simulation \virgopen SPH2\virgclose \ of P12, the final separation we obtain is $21$~\rs \ at $180$ days (the end of the dynamical in-spiral as defined above),  $16$~\rs \ at $1000$~days and $14$~\rs \ at the end of the simulation at $1500$ days (blue line in Figure~\ref{sph_separations}). The first two values can be compared to $21$~\rs \ at the end of the in-spiral and $18$~\rs \ at 1000 days for simulation \virgopen SPH2\virgclose \ of P12. We therefore find a very good level of agreement in the final separation obtained between the two codes. The small differences are mainly due to the differences in resolution and the sightly different initial separation of $100$~\rs \ that we had to adopt because the relaxed star in {\sc phantom} has a larger radius ($R=93$~\rs; defined using the volume-equivalent definition of \citealt{Nandez2014}) compared to the radius of the star stabilised in simulation \virgopen SPH2\virgclose \ of P12 ($R=83$~\rs).  

The second {\sc phantom} simulation, carried out with an initial separation of $218$~\rs, reaches a final separation of $29$~\rs, using the criterion to define the end of the rapid in-spiral described above or $22$~\rs\ at the end of the simulation ($5050$~days). We can compare these values ($29$ and $22$~\rs) to those obtained with {\sc phantom} in an initial binary configuration similar to that used by P12 ($21$~\rs \ and $16$~\rs). The visual comparison is shown in Figure~\ref{sph_separations}, where we plot the evolution of the separations of our two {\sc phantom} simulations by shifting the simulation starting at $100$~\rs \ by $3096$ days to a time when the other simulation, starting at $218$~\rs\ has a separation of $100$~\rs. The two {\sc phantom} simulations show final orbital separations that differ by $38$ per cent, corroborating the conclusion drawn from comparing the two {\sc enzo} simulations that the final orbital separation increases by including phases before the fast in-spiral.

\begin{table*}
\begin{center}
\begin{adjustbox}{max width=\textwidth}
\begin{tabular}{ccccccccccccccl}
\hline
$M_1$ & $M_{1,c}$ &$R_1$& Giant&$M_2$ & $q$ & $a_i/R_1$ & $\Omega / \omega^1$ &$R_1/R_{1,RL}$ & Code$^2$ & Resolution & $\tau_{\rm run}^3$& $a_f^4$ & $M_{\rm Unb}^5$ &Ref.$^6$\\
(\ms)&(\ms)&(\rs)&&(\ms)&&&&&&(Part./\rs)&(day)&(\rs)&(\%)&\\
\hline
4     & 0.7 & 66   & RGB & 0.7 & 0.18 & 1.6 & 1(y)     & 1.3 &SPH&500k&124&1(e)&10(?)&1\\
3     & 0.7 & 200 & AGB & 0.4 & 0.13 & 1.4 & 1(n)     & 1.3 &n-grid&2.2&800&4.4(a)&41(?)&2\\
3     & 0.7 & 200 & AGB & 0.4 & 0.13 & 1.4 & 0         & 1.3 &n-grid&2.2&800&4.7(a)&46(?)&2\\
5     & 1.0 & 200 & AGB & 0.4 & 0.08 & 1.4 & 1(n)     & 1.2 &n-grid&2.2&800&4.4(a)&21(?)&2\\
5     & 1.0 & 200 & AGB & 0.6 & 0.12 & 1.4 & 1(n)     & 1.3 &n-grid&2.2&800&4.8(a)&45(?)&2\\
5     & 0.94 &354& AGB & 0.6 & 0.12 & 1.5 & 0          & 1.2 &n-grid&2.2&800&8.9(a)&46(?)&2\\

2 & 0.335 & 44 & RGB & 0.2 & 0.1 & 1.3 & 0.14(n) & 3.7 & n-grid & 2.2 & 1050 & 2.1(a) & 3(?) & 3 \\
1 & 0.28 & 22 & RGB & 0.35 & 0.35 & 1.3 & 0.23(n) & 2.6 & n-grid & 2.2 & 1050 & 1.8(a) & 10(?) & 3 \\
1 & 0.45 & 243 & RGB & 0.35 & 0.35 & 1.3 & 0.24(n) & 2.6 & n-grid & 2.2 & 1050 & 21(a) & 11(?) & 3 \\
1 & 0.45 & 221 & RGB & 0.1 & 0.1 & 1.3 & 0.14(n) & 3.7 & n-grid & 2.2 & 1050 & 33(a) & 4(?) & 3 \\
2 & 0.45 & 177 & RGB & 0.35 & 0.18 & 1.3 & 0.18(n) & 3.2 & n-grid & 2.2 & 1050 & 19(a) & 6(?) & 3 \\
1 & 0.28 & 18 & RGB & 0.45 & 0.45 & 1.3 & 0.26(n) & 2.5 & n-grid & 2.2 & 1050 & 2.4(a) & 14(?) & 3 \\

1.05& 0.36 &31  & RGB & 0.6 & 0.57 & 2.0 & 0.95(n)&1.2 &a-grid(F)&0.29&60&9(e)&26(t)&4\\

0.88 & 0.39 & 85 & RGB & 0.1   & 0.11 &1 .0&0 & 1.8 &u-grid(E)&1.7&1000&   5.7(a) / 4.2(e)  &--&5\\
0.88 & 0.39 & 85 & RGB & 0.15 & 0.17 & 1.0&0 & 1.9 &u-grid&(E)1.7&1000& 6.9(a) / 4.7(e)  &--&5\\
0.88 & 0.39 & 85 & RGB & 0.3   & 0.34 & 1.0&0 &  2.1 &u-grid(E)&1.7&1000&   11(a) /  9.0(e)  &--&5\\
0.88 & 0.39 & 85 & RGB & 0.6   & 0.68 & 1.0&0 & 2.4 &u-grid(E)&1.7&1000&   19(a) / 16(e)    &--&5\\
0.88 & 0.39 & 85 & RGB & 0.9   & 1.02 & 1.0&0 & 2.6 &u-grid(E)&1.7&1000&   26(a) /  22(e)   &--&5\\

0.88&0.39&83&RGB&0.1&0.11&1.0&0&1.8 &SPH(S)&500k&1050&  6.1(a) /  5.7(e) &2(t)&5\\
0.88&0.39&83&RGB&0.15&0.17&1.0&0&1.9 &SPH(S)&500k&950&  7.3(a) /  7.8(e) &6(t)&5\\
0.88&0.39&83&RGB&0.3&0.34&1.0&0&2.1 &SPH(S)&500k&750&    11(a)  /  10(e)  &8(t)&5\\
0.88&0.39&83&RGB&0.6&0.68&1.0&0&2.4&SPH(S)&500k&950&    21(a)  /  18(e)  &10(t)&5\\
0.88&0.39&83&RGB&0.9&1.02&1.0&0&2.6 &SPH(S)&500k&600&    27(a)  /  25(e)  &10(t)&5\\

1.50&0.32&26$^7$&RGB&0.36&0.24&2.0&0&1.0&SPH(SM)&200k&(?)&0.91(e)&100(r)$^8$&6\\

1.98&0.38&49&RGB&0.99&0.5&1.0&0.95(n)&2.3&m-mesh(A)&0.07-0.01&120&4.9(e)&8(t)&7\\

1.20&0.32&29&RGB&0.32&0.27&2.0&0&1&SPH(SM)&100k&2000&1.4&100&8\\
1.20&0.32&29&RGB&0.36&0.30&2.1&0&1&SPH(SM)&100k&2000&1.5&100&8\\
1.20&0.32&29&RGB&0.40&0.33&2.1&0&1&SPH(SM)&100k&2000&1.4&100&8\\
1.40&0.32&28&RGB&0.32&0.23&2.0&0&1&SPH(SM)&100k&2000&1.1&100&8\\
1.40&0.32&28&RGB&0.36&0.26&2.0&0&1&SPH(SM)&100k&2000&1.1&100&8\\
1.40&0.32&28&RGB&0.40&0.29&2.0&0&1&SPH(SM)&100k&2000&1.2&100&8\\
1.60&0.32&26&RGB&0.32&0.20&1.9&0&1&SPH(SM)&100k&2000&0.87&100&8\\
1.60&0.32&26&RGB&0.36&0.23&2.0&0&1&SPH(SM)&100k&2000&0.91&100&8\\
1.60&0.32&31&RGB&0.36&0.23&1.6&1(y)&1&SPH(SM)&100k&2000&0.93&100&8\\
1.60&0.32&26&RGB&0.40&0.25&2.0&0&1&SPH(SM)&100k&2000&0.96&100&8\\
1.80&0.32&16&RGB&0.32&0.18&1.9&0&1&SPH(SM)&100k&2000&0.43&100&8\\
1.80&0.32&16&RGB&0.36&0.20&1.9&0&1&SPH(SM)&100k&2000&0.48&100&8\\
1.80&0.32&16&RGB&0.40&0.22&2.0&0&1&SPH(SM)&100k&2000&0.53&100&8\\
1.18&0.36&60&RGB&0.32&0.27&2.0&0&1&SPH(SM)&100k&2000&3.2&100&8\\
1.18&0.36&60&RGB&0.36&0.31&2.1&0&1&SPH(SM)&100k&2000&3.7&100&8\\
1.18&0.36&60&RGB&0.40&0.34&2.1&0&1&SPH(SM)&100k&2000&3.5&100&8\\
1.38&0.36&57&RGB&0.32&0.23&2.0&0&1&SPH(SM)&100k&2000&2.5&100&8\\
1.38&0.36&57&RGB&0.36&0.26&2.0&0&1&SPH(SM)&100k&2000&2.8&100&8\\
1.38&0.36&57&RGB&0.40&0.29&2.0&0&1&SPH(SM)&100k&2000&3.0&100&8\\
1.59&0.36&50&RGB&0.32&0.20&1.9&0&1&SPH(SM)&100k&2000&1.7&100&8\\
1.59&0.36&50&RGB&0.36&0.23&2.0&0&1&SPH(SM)&100k&2000&1.8&100&8\\
1.59&0.36&50&RGB&0.40&0.25&2.0&0&1&SPH(SM)&100k&2000&2.1&100&8\\
1.80&0.36&41&RGB&0.32&0.18&1.9&0&1&SPH(SM)&100k&2000&1.2&100&8\\
1.80&0.36&41&RGB&0.36&0.20&1.9&0&1&SPH(SM)&100k&2000&1.3&100&8\\
1.80&0.36&41&RGB&0.40&0.22&2.0&0&1&SPH(SM)&100k&2000&1.4&100&8\\

0.88&0.39&100&RGB&0.6&0.68&3&0&0.81&a-grid(E)&1.7&2000&36(a)/20(e)&16(t)&9\\
0.88&0.39&93&RGB&0.6&0.68&1.1&0&2.2&SPH(P)&1m&1850&21(a)/16(e)&13(t)&9\\
0.88&0.39&91&RGB&0.6&0.68&2.4&0&1.0&SPH(P)&300k&5050&29(a)/22(e)&16(t)&9\\

\hline
\multicolumn{15}{l}{$^1$Stellar spin frequency as a function of orbital frequency, with an indication of whether the star was stabilised in its rotating configuration (y) or not (n)} \\
\multicolumn{15}{l}{before the start of the simulation.}\\
\multicolumn{15}{l}{$^2$SPH: smoothed particle hydrodynamics; u-grid: uniform, static grid; n-grid: static nested grids; m-mesh: moving mesh; a-grid: adaptive }\\
\multicolumn{15}{l}{mesh refinement grid; F: {\sc flash}, E: {\sc Enzo}, S: {\sc snsph}, P: {\sc phantom}, SM: {\sc starsmasher}. A: {\sc arepo}}\\
\multicolumn{15}{l}{$^3$ Information not provided (?).}\\
\multicolumn{15}{l}{$^4$Rounded to 2 significant figures, calculated either at the end of the simulation (e) or at a time defined by the formula in Section \ref{ssec:separation} (a).}\\
\multicolumn{15}{l}{$^5$ Calculated by including thermal energy (t), not including thermal energy (k), information not provided (?) or including recombination energy (r).}\\
\multicolumn{15}{l}{$^6$1: \citealt{Rasio1996}. 2: \citealt{Sandquist1998}. 3: \citealt{Sandquist2000}. 4: \citealt{Ricker2012}. 5: \citealt{Passy2012}. 6: \citealt{Nandez2015}.}\\
\multicolumn{15}{l}{7: \citealt{Ohlmann2016}. 8: \citealt{Nandez2016}. 9: This work.}\\
\multicolumn{15}{l}{$^7$ This is the Roche lobe radius also corresponding to the SPH radius in their simulation.}\\
\multicolumn{15}{l}{$^8$ Note that the same simulation run without recombination energy unbinds 50 per cent of the envelope, although the authors of that simulations}\\
\multicolumn{15}{l}{do not present data to illustrate their statement.}\\
\end{tabular}
\end{adjustbox}
\end{center}
 \begin{quote}
  \caption{\protect\footnotesize{ A comparison of initial conditions and final outcomes of previous common envelope simulations that included at least one giant star.}} \label{comparison_with_previous_works_data}
 \end{quote}
\end{table*}

\subsection{Envelope ejection}
\label{ssec:gas}

To determine the extent to which the envelope is unbound we determined whether gas has total energy larger than zero. The total energy can be calculated including or excluding thermal energy, where the former prescription results in more unbound gas. \citet{Ivanova2011} discussed how it is the enthalpy rather than the thermal energy that needs to be included when determining whether a gas parcel is bound or not. Using enthalpy instead of thermal energy increases the unbound mass very marginally. In this work, where not otherwise specified, we include thermal energy in the computation of the bound and unbound mass.

\begin{figure*}
\centering     
\subfigure[]{\label{fig:a}\includegraphics[scale=0.7, trim=6.5cm 0.0cm 5.5cm 0.0cm, clip]{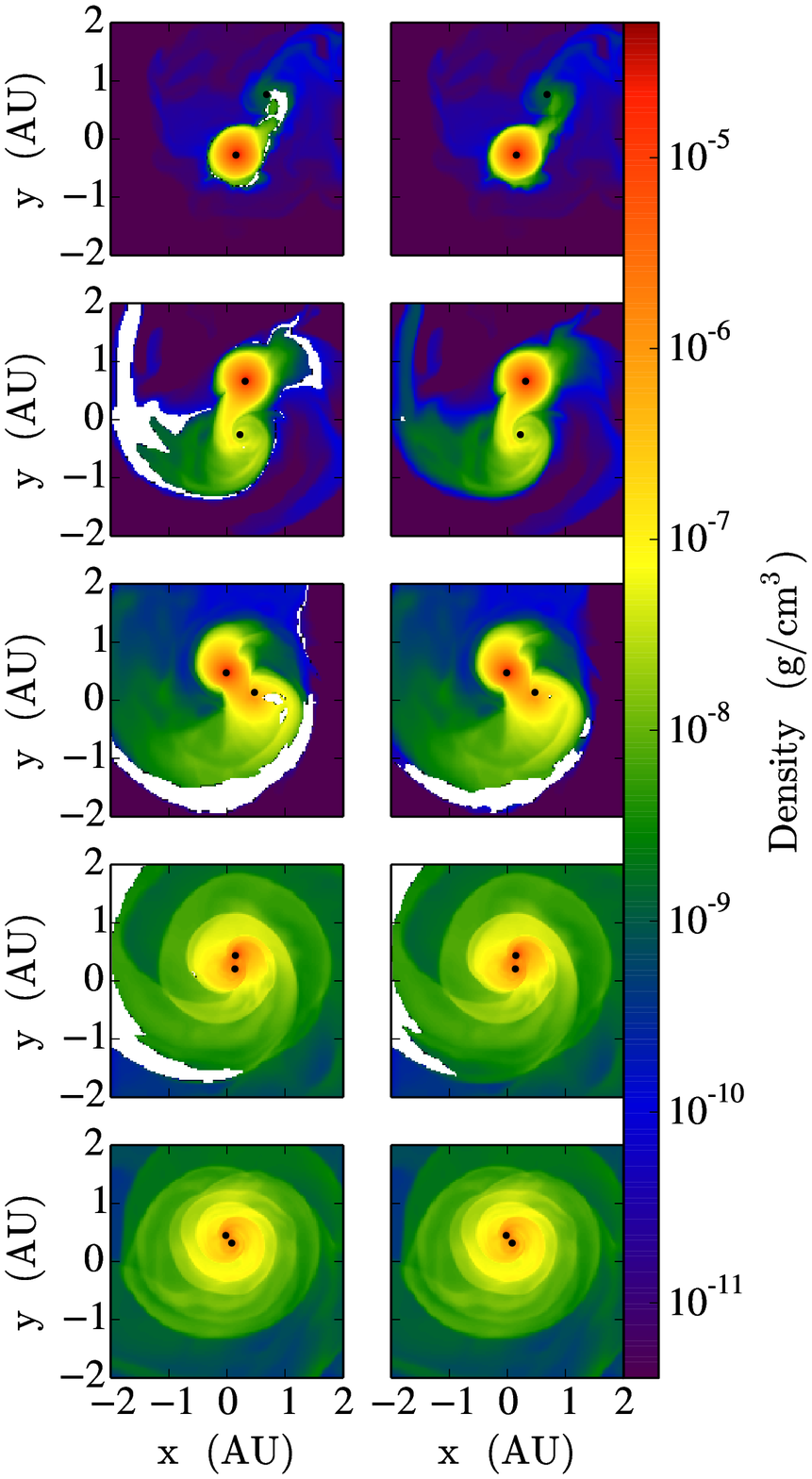}}
\subfigure[]{\label{fig:b}\includegraphics[scale=0.7, trim=9.cm 0.0cm 8.5cm 0.0cm, clip]{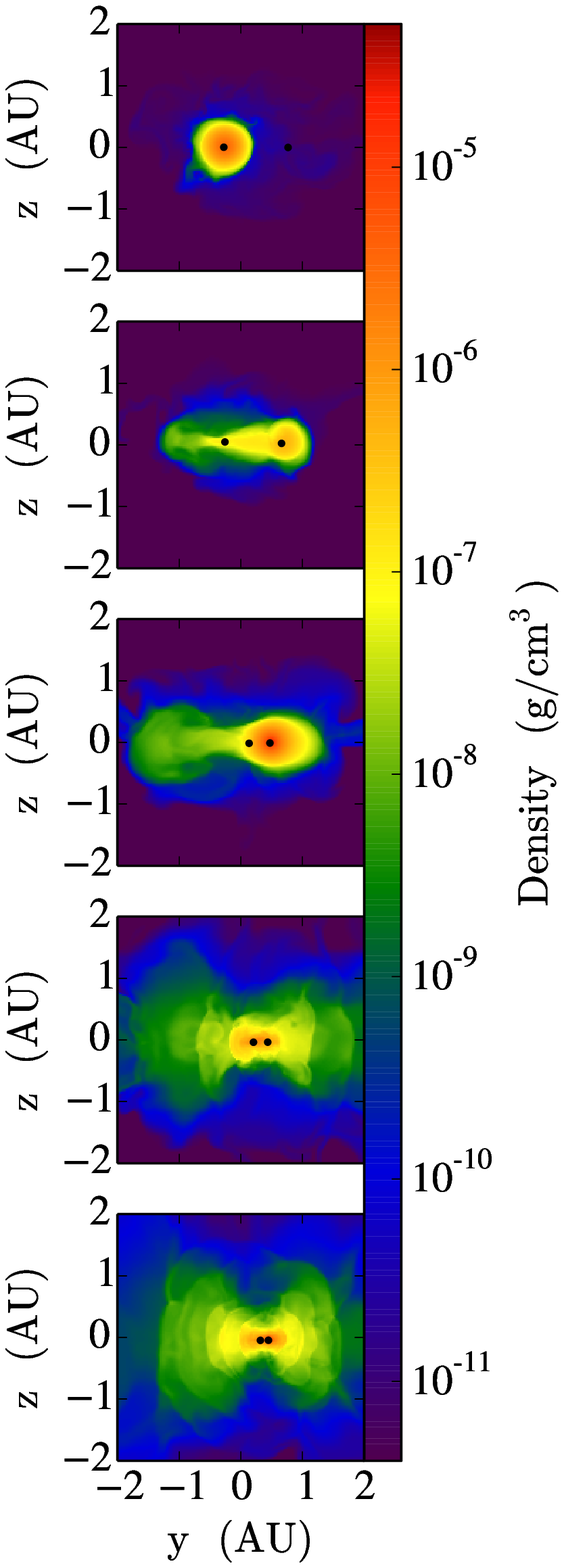}}
\caption{\protect\footnotesize{Left panel, left column: density slices perpendicular to the $z$ axis in the orbital plane after (from top to bottom) $887$, $1381$, $1653$, $1774$ and $1840$~days from the beginning of the {\sc Enzo} simulation. The point-mass particles representing the core of the primary and the companion are shown as black dots, while the white regions represent the unbound gas. The size of the black dots is not representative of any property of the point masses and is chosen only to highlight them. Left panel, right column: same as the left column, but excluding thermal energy ($E_{\rm th}$) in the computation of the bound/unbound mass elements. Right panel: density slices perpendicular to the orbital plane at $x=0$, taken at the same times as the left panels.}}
\label{bound_unbound_z_slices}
\end{figure*}

For the {\sc Enzo} simulation we present the density slices in the orbital and perpendicular planes in Figure~\ref{bound_unbound_z_slices}. In the first and middle columns we compare the distribution of unbound gas both using thermal energy (left column) and not (middle column), to distinguish between gas acceleration and gas heating. The initial unbinding event (first two rows, left columns) happens because of heating of the gas falling into the potential well of the companion during the mass transfer phase, which is why this unbound material is not recorded on Figure~\ref{bound_unbound_z_slices}, middle column. This unbound material has very low mass. Later, during the rapid in-spiral phase (Figure~\ref{bound_unbound_z_slices}, last three rows, left and middle columns) far more mass is unbound because it is accelerated above the escape velocity as demonstrated by the similarity of the left and central columns.

\begin{figure*}
\centering     
\subfigure[]{\label{fig:a}\includegraphics[scale=0.7, trim=9cm 0.0cm 9cm 0.0cm, clip]{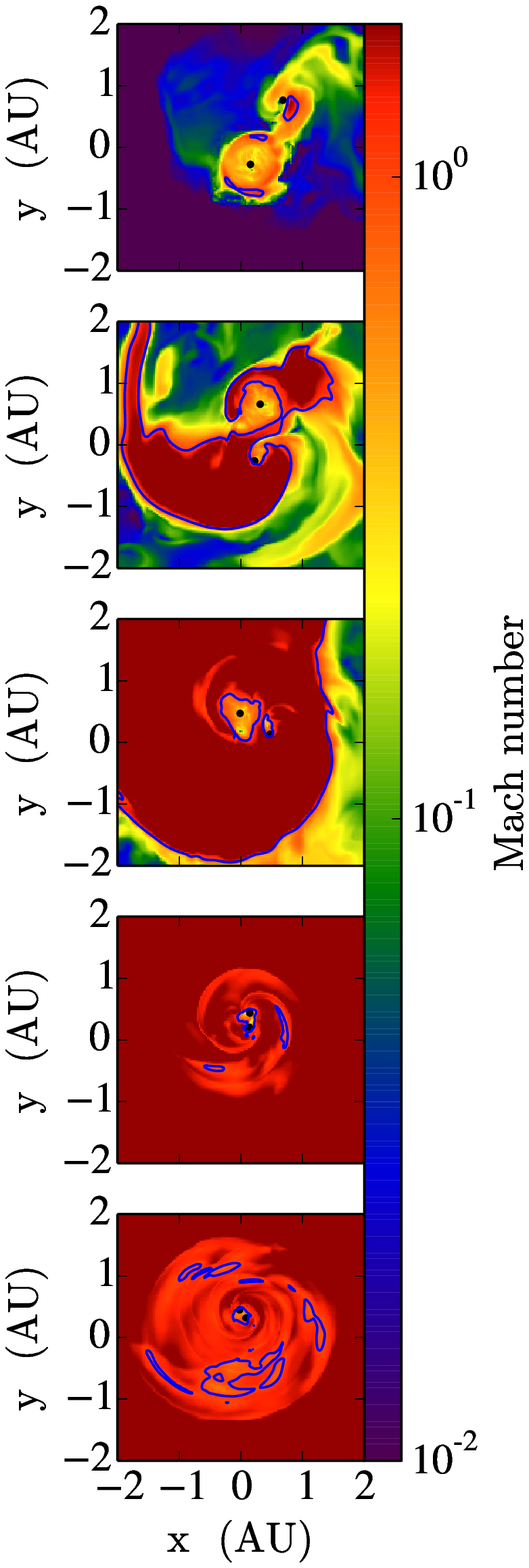}}
\subfigure[]{\label{fig:b}\includegraphics[scale=0.7, trim=9cm 0.0cm 8cm 0.0cm, clip]{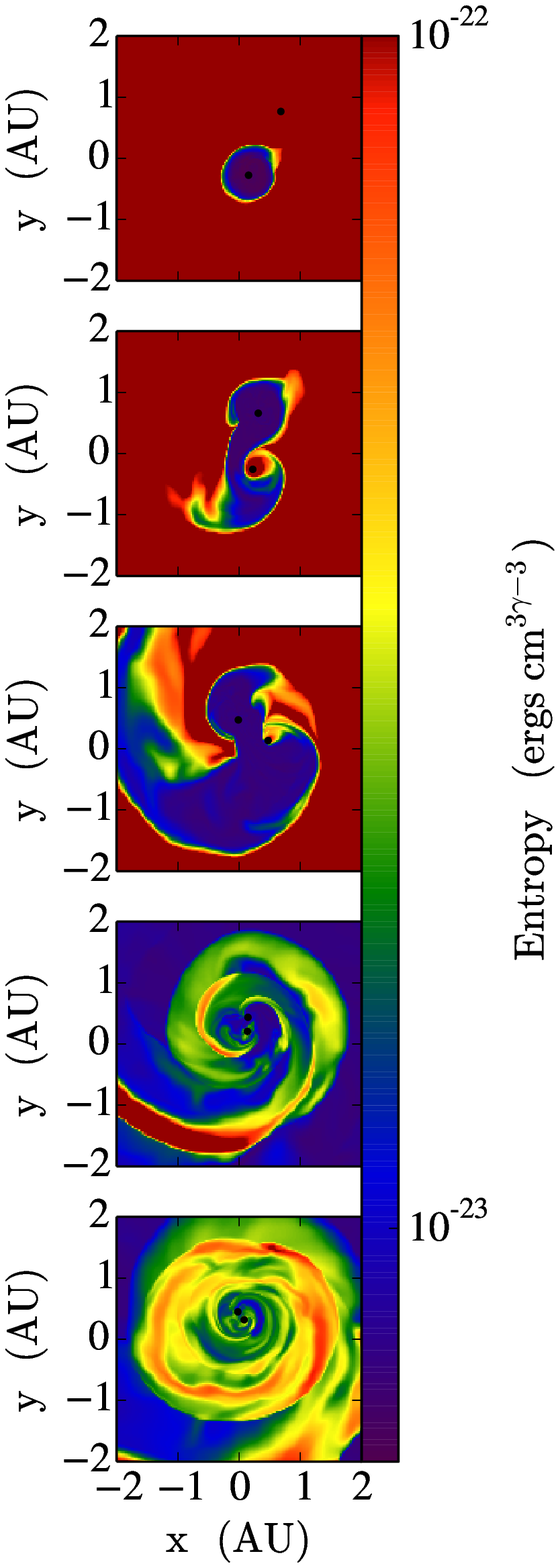}}
\caption{\protect\footnotesize{Left panel: Mach number slices perpendicular to the $z$ axis in the orbital plane for the {\sc Enzo} simulations, after (from top to bottom) $887$, $1381$, $1653$, $1774$ and $1840$~days from the beginning of the simulation. The point-mass particles representing the core of the primary and the companion are shown as black dots. The size of the black dots is not representative of any property of the point masses and is chosen only to highlight them. The Mach number equal to unity contours are marked with a blue line. Right panel: same as for the right panel, but for the entropy distribution.}}
\label{machnumber_entropy_z_slices}
\end{figure*}

Similarly to what was reported in previous work (\citealt{Sandquist1998}, \citealt{Ricker2012}, \citealt{Ohlmann2016}), we observe that, while the pre-contact interactions do not accelerate the envelope gas to supersonic speeds, during the in-spiral a bow shock forms in front of the companion followed by spiral shocks generated both by primary's core and the companion. This behaviour is showed in Figure~\ref{machnumber_entropy_z_slices}, where we plot the envelope Mach number and the gas entropy in the orbital plane during the rapid in-spiral (lasting from $1515$~days, or $4.2$~yr, to $1840$~days, or $5$~yr, from the beginning of the simulation). The spiral shocks wind around the binary and are stronger closer to the point-masses, as highlighted by the entropy distribution in the last two panels of Figure~\ref{machnumber_entropy_z_slices} (right column). We note that the high entropy in the peripheral regions in the first two slices of Figure~\ref{machnumber_entropy_z_slices} (right column) are due to residual \virgopen vacuum\virgclose \  gas with very high temperature.

The evolution of the unbound gas can be followed only inside the simulation box, due to the grid nature of {\sc Enzo}. However, we estimated whether the mass that leaves the box is bound or unbound in the following way. We calculated the fraction of unbound gas contained within the box boundary (i.e., within the six, one cell thick, box faces) and we assumed it to be representative of the fraction of unbound gas between code outputs (which take place every $3.65$~days~$=0.01$~yr). We then multiplied this fraction by the mass that leaves the box between code outputs.

\begin{figure}
\centering     
\subfigure[]{\label{fig:a}\includegraphics[scale=0.4]{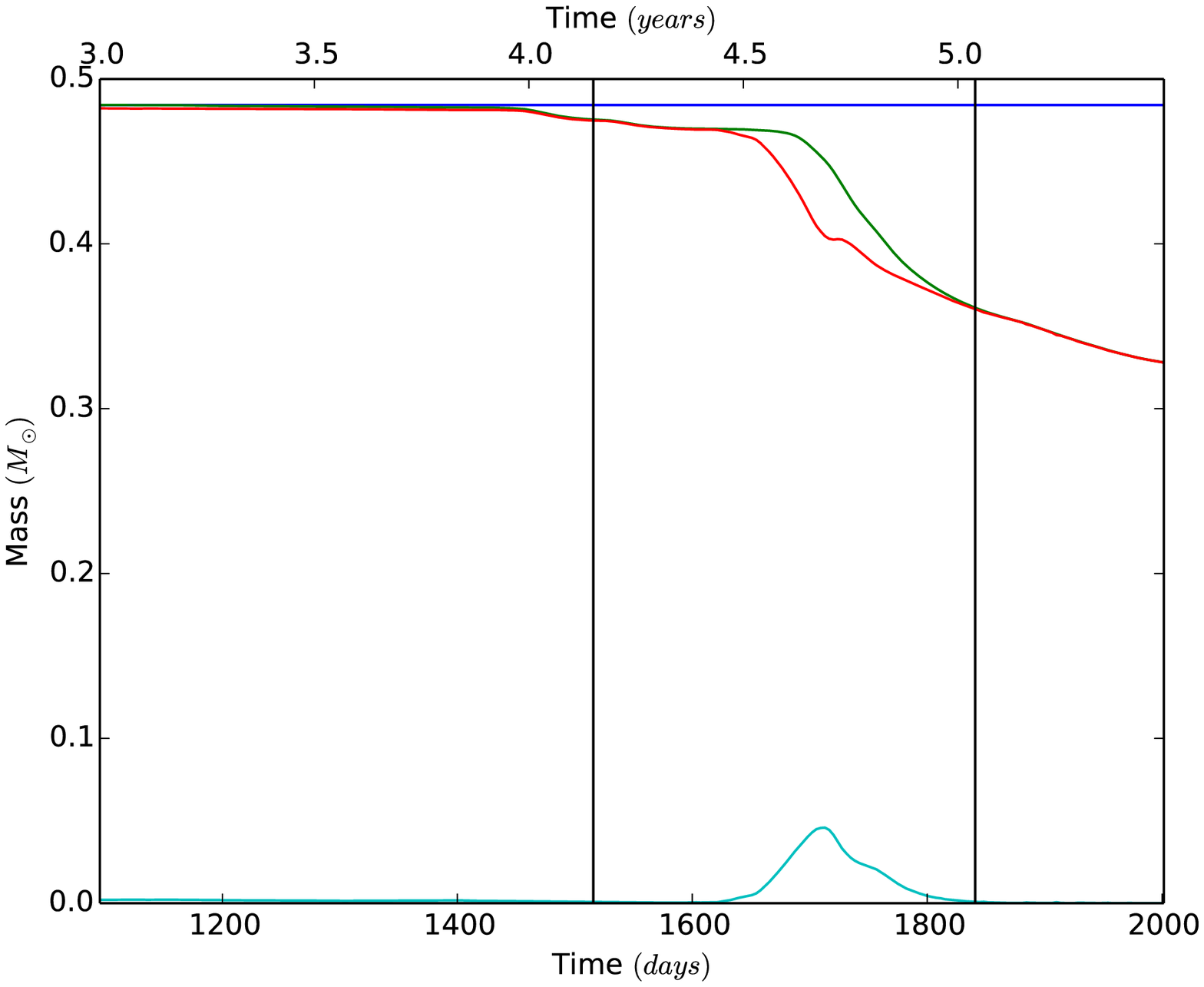}}
\subfigure[]{\label{fig:b}\includegraphics[scale=0.4]{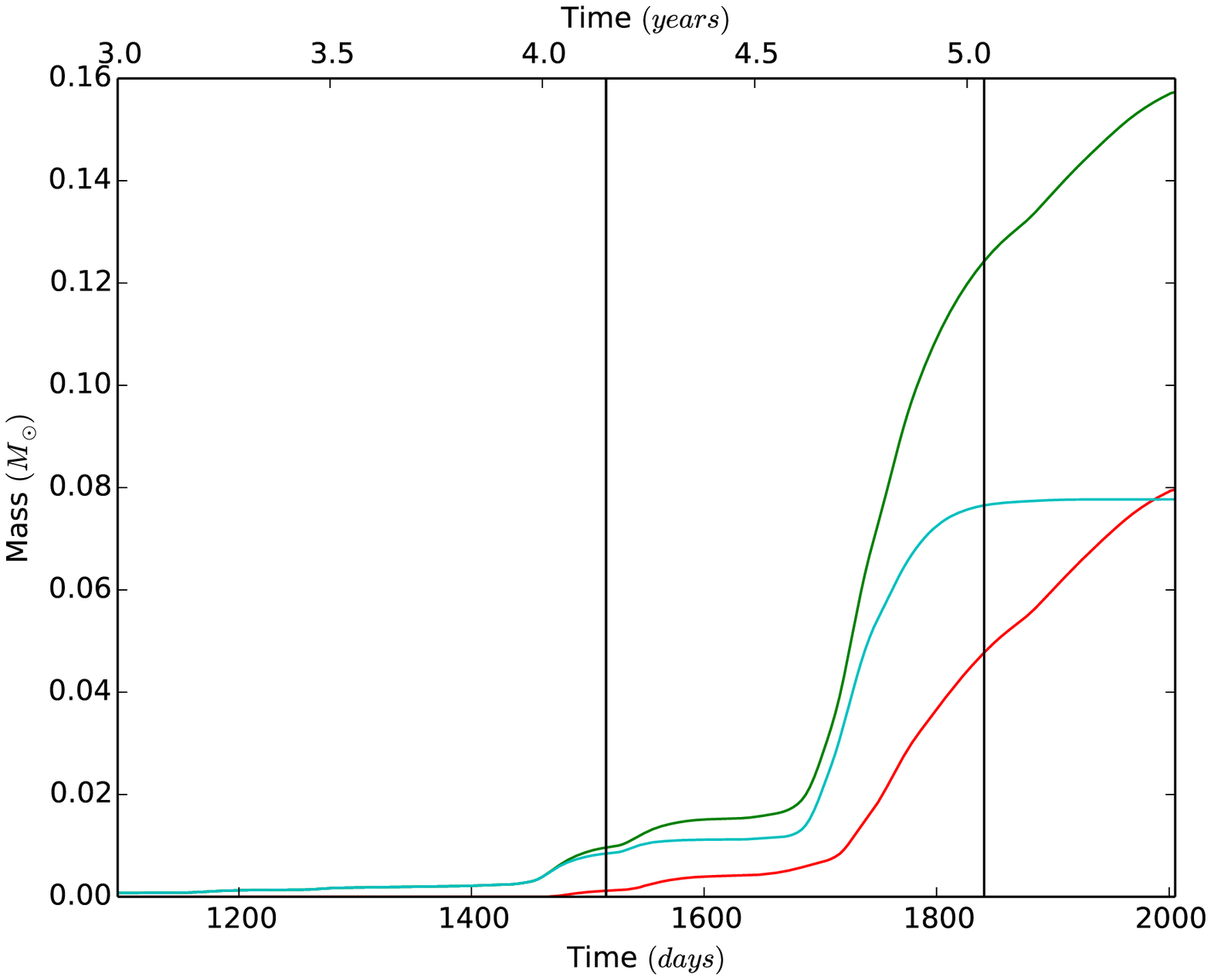}}
\caption{\protect\footnotesize{Upper panel: evolution of the gas mass inside the {\sc Enzo} simulation domain over time. The blue line represents the value of the initial gas mass contained in the domain and is plotted for comparison, while the green line shows the evolution of the total mass contained inside the box. The red and cyan lines show, respectively, the bound and unbound components of the mass. Lower panel: cumulative mass of the gas flowing out of the simulation box over time for the same simulation. Line colours have the same meaning as for the upper panel. The black vertical lines in both panels correspond to the beginning and end of the rapid in-spiral and both the plots are limited to the part of the simulation where significant mass is lost from the box.}}
\label{mass_boundness}
\end{figure}

The estimate of the total unbound mass leaving the box is shown in Figure~\ref{mass_boundness} (lower panel). Our approximation is consistent with the total amount of mass that leaves the box during the simulation, shown in Figure \ref{mass_boundness} (upper panel). The first unbound mass leaves the box at approximately 1500~days, at the onset of the rapid in-spiral, but the bulk of the mass flows out during the rapid in-spiral phase (between approximately 1750 and 1900~days). The total mass unbound in the simulation amounts to $8 \times 10^{-2}$~\ms, or $16$ per cent of the initial envelope mass. The unbound mass is $14$ per cent, if we do not include thermal energy and $17$ per cent, if we use the enthalpy as suggested by \citet{Ivanova2011}. P12 found that $10$ per cent of the initial envelope mass was unbound, which should be compared to our $16$ per cent. This increase likely represents the effect of a larger initial separation.

\begin{figure}
\includegraphics[scale=0.45, trim=0.8cm 0.0cm 0.0cm 0.0cm]{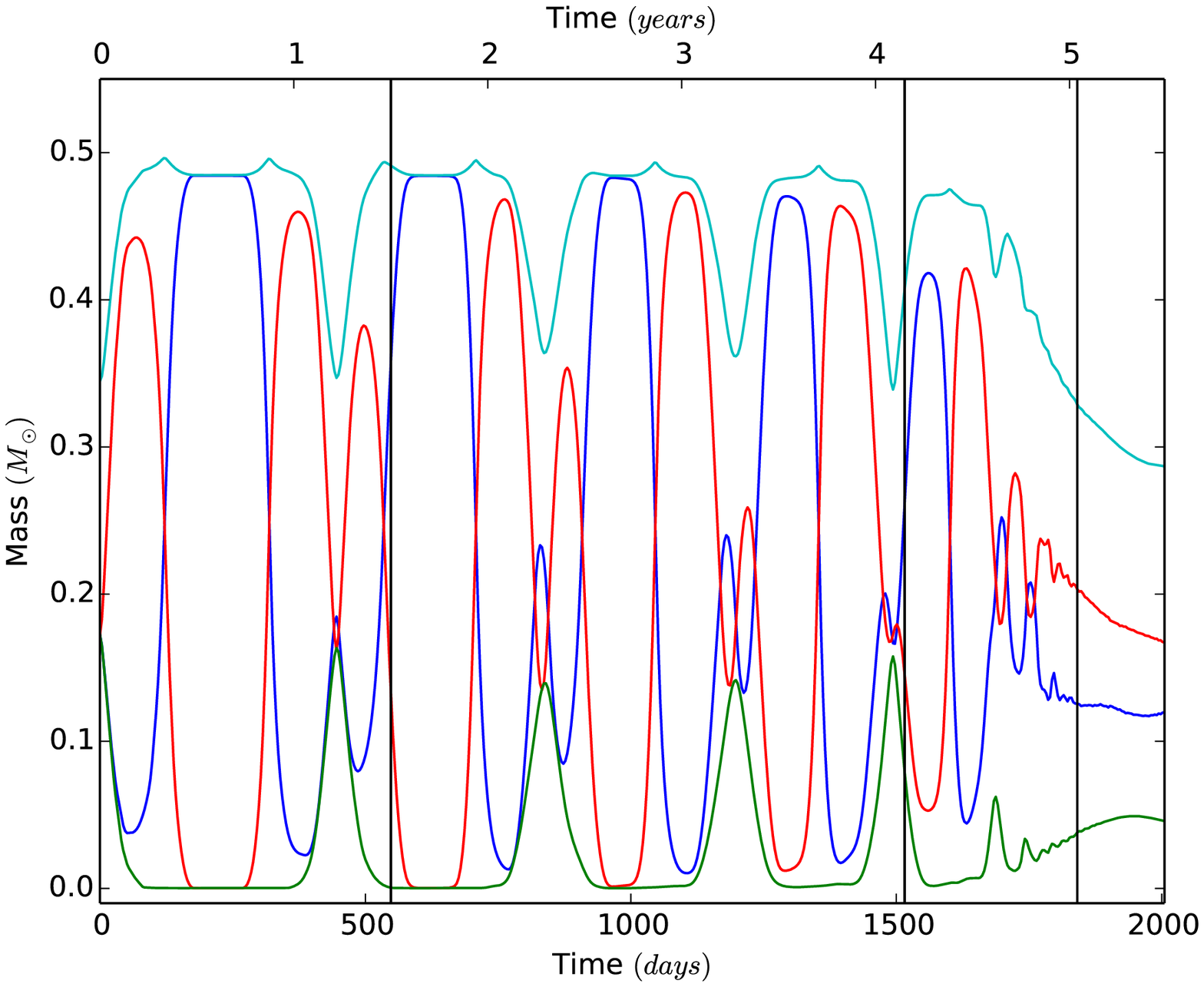}
\caption{\protect\footnotesize{Gas mass inside the simulation domain vs time for the gas located in six pyramids whose bases are the six faces and whose vertexes  are at the centre of the domain, for the {\sc Enzo} simulation. The two pyramids along the $x$ axis are in blue, along the $y$ axis are in red and along the $z$ axis are in green. The cyan line shows the sum of the $x$ and $y$ contributions to highlight the behaviour of the mass ejection in the orbital plane. The black vertical lines show the estimated beginning of the Roche lobe overflow phase and the beginning and end of the rapid in-spiral phase.}}
\label{ejecta_geometry}
\end{figure}

Most of the ejecta is expected to flow away close to the orbital plane, where the gas is accelerated by the orbiting particles. This was already borne out by the simulations of \citet{Sandquist1998} and is clearly seen in Figure~\ref{bound_unbound_z_slices} (right panel).
Figure~\ref{ejecta_geometry} demonstrates how the envelope is ejected around the binary over time. We divide the computational domain into six pyramids centred at the centre of the box and whose bases are the six faces. We plot the mass in pairs of pyramids aligned with each of the three directions, $x$, $y$ and $z$. Initially the mass is equally distributed in the three pairs of pyramids as the star resides at the centre of the box. Later the mass in the pyramid pairs oscillates as the giant moves along its orbit. The decrease of the peaks in the green line during the rapid in-spiral phase in Figure~\ref{ejecta_geometry} marking approximately the completion of a full orbital revolution, demonstrates a decrease in the mass contained in the $z$ direction in favour of mass contained in the other two directions. The decreasing amplitude of the oscillations over time indicates that the gas distribution becomes more and more independent of the orbital motion of the two particles, as the interaction proceeds. Towards the end of the CE, as the oscillations cease, more mass is being ejected out of the simulation box highlighting how the rapid in-spiral rapidly lifts the envelope, disrupting the primary star.

\begin{figure}
\centering     
\subfigure[]{\label{fig:a}\includegraphics[scale=0.4]{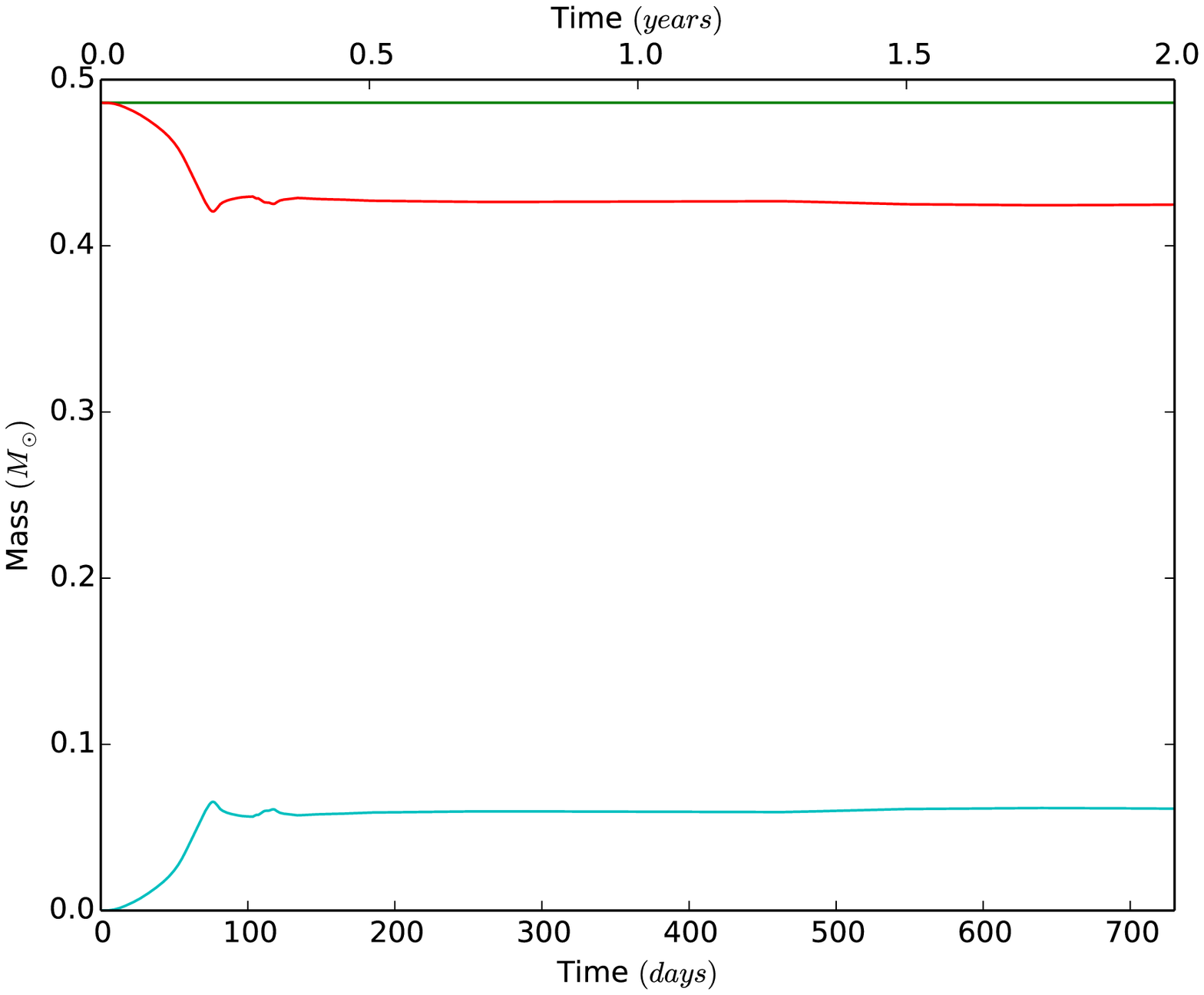}}
\subfigure[]{\label{fig:b}\includegraphics[scale=0.4]{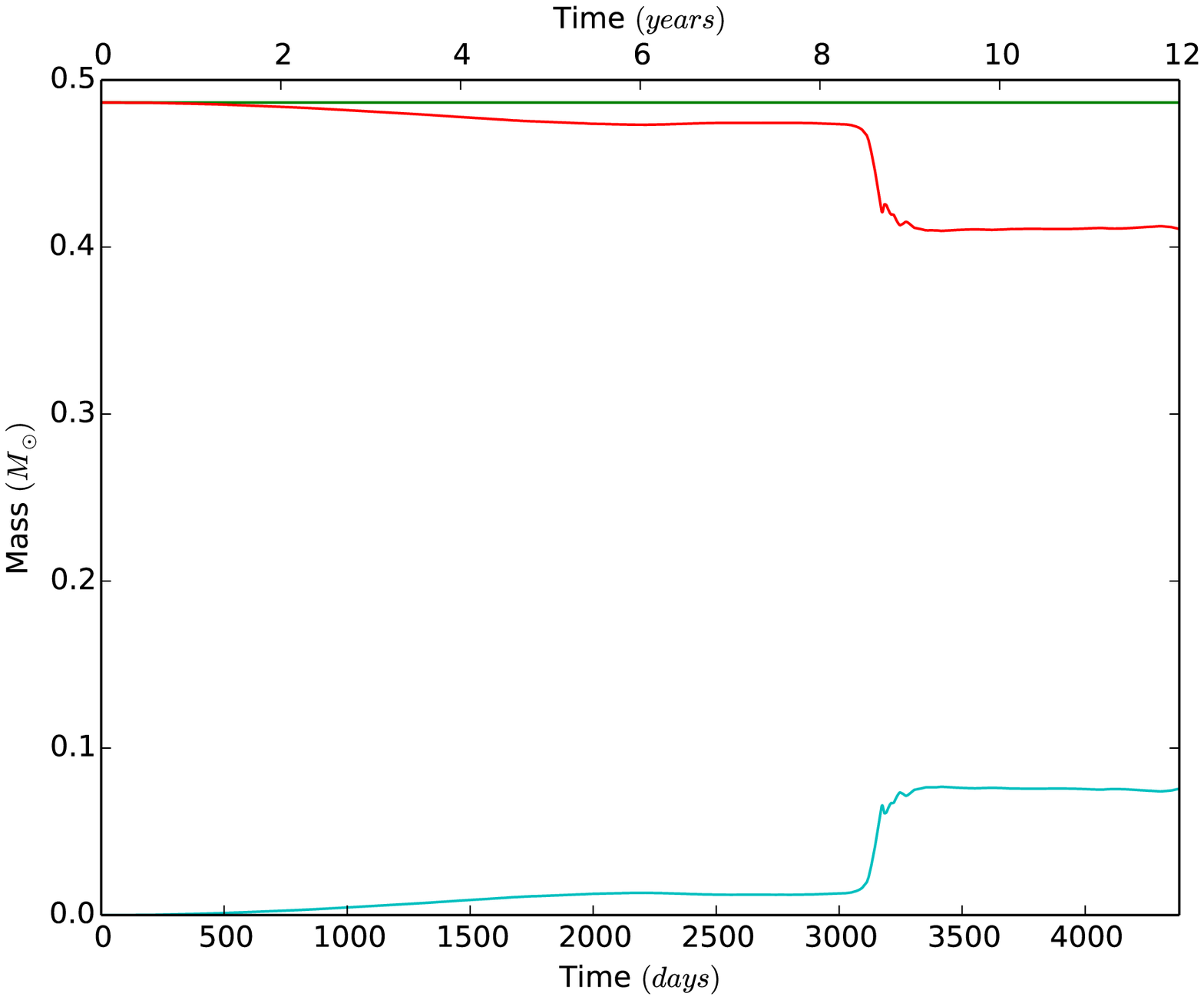}}
\caption{\protect\footnotesize{Upper panel: evolution of the gas mass for the {\sc phantom} simulation starting with an initial separation of $100$~\rs. The green line shows the evolution of the total mass. The red and cyan lines show, respectively, the bound and unbound components of the mass. Lower panel: evolution of the gas mass for the {\sc phantom} simulation starting with an initial separation of $218$~\rs. Colours are the same as those of the upper panel.}}
\label{mass_boundness_sph}
\end{figure}

For the {\sc phantom} simulations, we plot the evolution of the bound and unbound components of the mass in Figure~\ref{mass_boundness_sph}. The first simulation, starting from an initial orbital separation of $100$~\rs, shows that the unbinding of the envelope mass begins right after the  simulation is started and terminates around $80$~days, before the separation has levelled off (Figure~\ref{mass_boundness_sph}, upper panel). The unbinding is almost entirely caused by gas accelerated above escape velocity, while heating plays a minor role. Both these results are in agreement to what was obtained by P12's \virgopen SPH2\virgclose \ simulation. The mass unbound is approximately 13 per cent of the envelope mass, compared to approximately $10$ per cent for \virgopen SPH2\virgclose \ of P12. Again, we think that these differences are due to differences in the code used and in the slightly larger initial separation.

The second {\sc phantom} simulation starts instead from an initial orbital separation of $218$~\rs. In this case the unbinding of the mass begins gradually while the companion approaches the primary star, but before the onset of the rapid in-spiral. Then as soon as the rapid in-spiral is triggered the bulk of the mass is ejected and unbound (Figure~\ref{mass_boundness_sph}, lower panel). This behaviour is very similar to what we obtained for our {\sc Enzo} simulation starting from an initial separation of $300$~\rs. The mass unbound in the simulation is $16$ per cent of the envelope mass, marginally larger than the $13$ per cent for the same simulation starting with a lower initial separation. The increased amount of mass unbound is therefore in line with what we obtained for our {\sc Enzo} simulations. Similarly to what was observed in the {\sc Enzo} simulation, and in line with previous work, the envelope is mainly expelled in the orbital plane and a series of spiral shocks are produced while primary core and companion in-spiral towards each other. 

\subsection{Tidal bulges}
\label{ssec:bulges}

As explained in Section~\ref{ssec:binary}, the pre-contact phase in our simulation takes place over much shorter time-scales than it would in nature. The short pre-contact time-scale observed in our simulation is due to deformations created on the primary by the insertion of the companion into the computational domain. This is likely the result of the lack of stabilisation of the binary in the co-rotating frame, discussed in Section~\ref{ssec:binary}. 

A  simple order of magnitude analytical estimate of the mass, $\delta M_1$, contained in the tidal bulges of the primary, for equilibrium tides, can be obtained from \citet{Zahn2008}:

\begin{equation}
\delta M_1 \leq M_2\Big( \frac{R_1}{a} \Big)^3 \ ,
\label{bulges_mass}
\end{equation}

\noindent where $M_1$, $M_2$, $R_1$ and $a$ are the masses of the primary, secondary, the radius of the primary and the orbital separation, respectively. For the purpose of this calculation we only vary $a$ with time, while leaving $R_1$ constant, hence the value of $\delta M_1$ oscillates due to the eccentricity that develops (Figure~\ref{separation_vs_time}, upper panel). 

\begin{figure}
\includegraphics[scale=0.45, trim=0.8cm 0.0cm 0.0cm 0.0cm]{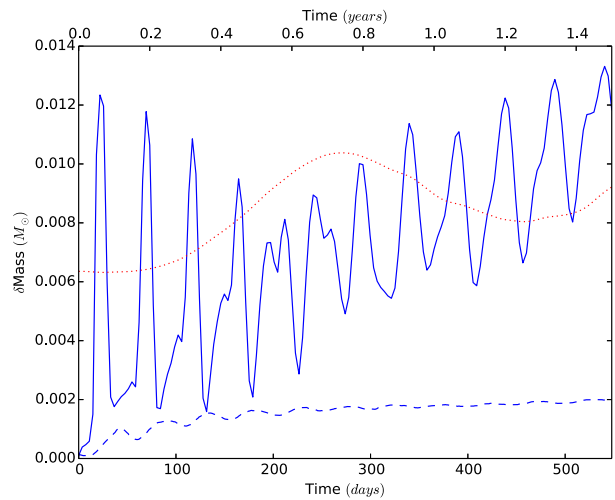}
\caption{\protect\footnotesize{Mass in the tidal bulges of the primary star overtime during the pre-contact phase, estimated from the {\sc enzo} simulation data (solid blue line), from the {\sc phantom} simulation with an initial separation of $218$~\rs \ data (dashed blue line) and from the analytical formula (dotted red line).}}
\label{tidal_bulge_mass}
\end{figure}

In Figure~\ref{tidal_bulge_mass} we compare this analytical estimate with the bound mass residing outside the initial equilibrium radius of the primary for {\sc enzo} (solid line) and {\sc phantom} (dashed line). The insertion of the companion into the simulations' domain triggers some oscillations on a time-scale of the order of the dynamical time of the star ($\simeq$21 days). Over the pre-contact phase there is also a gradual expansion of the star, seen as an increasing trend of the mass outside its original volume. The oscillation is caused by the mass distribution acquiring two small opposite bulges that are initially aligned with the direction of the companion, but which then disappear and reappear at 90 degrees to the original direction. This generates the relatively strong torques that contribute to the fast decrease of the orbital separation during the pre-contact phase.

Both simulations show that the bulge mass is commensurate with the analytical approximation. The oscillation is smaller in {\sc phantom} due to the fact that SPH is more stable to surface deformations compared to the grid-based {\sc Enzo}. This may be the reason why the tidal in-spiral is slower in {\sc phantom} than {\sc enzo}. Another reason, discussed more in depth in Sec.~\ref{ssec:conservation}, could be that {\sc enzo} conserves angular momentum less well than {\sc phantom}.

At $1515$~days in the {\sc enzo} simulation, gravitational drag between the companion and the surrounding gas becomes the main mechanism exchanging energy and causing the decrease of the orbital separation (see \citealt{Ricker2012} and discussion below in Section~\ref{ssec:density}). This regime change is evident in Figure~\ref{separation_vs_time}, at the location of the vertical line representing our estimation for the beginning of the rapid in-spiral.

\subsection{Evolution of the gas velocities and density in proximity to the companion}
\label{ssec:density}

The mechanism behind the energy and angular momentum exchange that drives the in-spiral is gravitational drag \citep{Ricker2012}. Gravitational drag is caused by the gas which flows past the moving body (in our case the companion star), forming a wake with higher density behind it that gravitationally pulls on it, slowing the body down. The gravitational drag experienced by a body immersed in a fluid depends on the body's mass, the fluid density, the velocity contrast between the body and the fluid and on the Mach number of the body. Approximations for the gravitational drag are given by \citet[][$F_{\mathrm{drag}} \propto (M_2 \rho v_{\mathrm{rel}}^2) / (v_{\mathrm{rel}}^2 + c_\mathrm{s}^2)^2$, for the subsonic motion regime]{Iben1993} and by \citet{Ostriker1999} who calculated a more detailed formula, carefully considering the effects of the Mach number. 

It is fundamental to determine whether simulations accurately reproduce the effects of gravitational drag because this determines in turn when the companion in-spiral terminates and, as a result, the amount of orbital energy deposited. Is the end of the in-spiral due to the decreasing density around the particles, the co-rotation of the surrounding gas or a change in the Mach regime (as was the case in the simulations of \citealt{Staff2016b})? Does the density gradient affect the force as questioned by \citet{MacLeod2015}? How does the interplay of resolution and smoothing length affect the simulation (\citealt{Staff2016a})? It is well known that the particles will not approach closer than approximately two smoothing lengths, effectively because their potentials are flat within that distance. However, less clear are the effects that not resolving a radius of the order of the Bondi radius \citep{Bondi1952} around the particles will have on the drag force (\citealt{Staff2016b}).

\begin{figure}
\centering     
\subfigure[]{\label{fig:b}\includegraphics[scale=0.4]{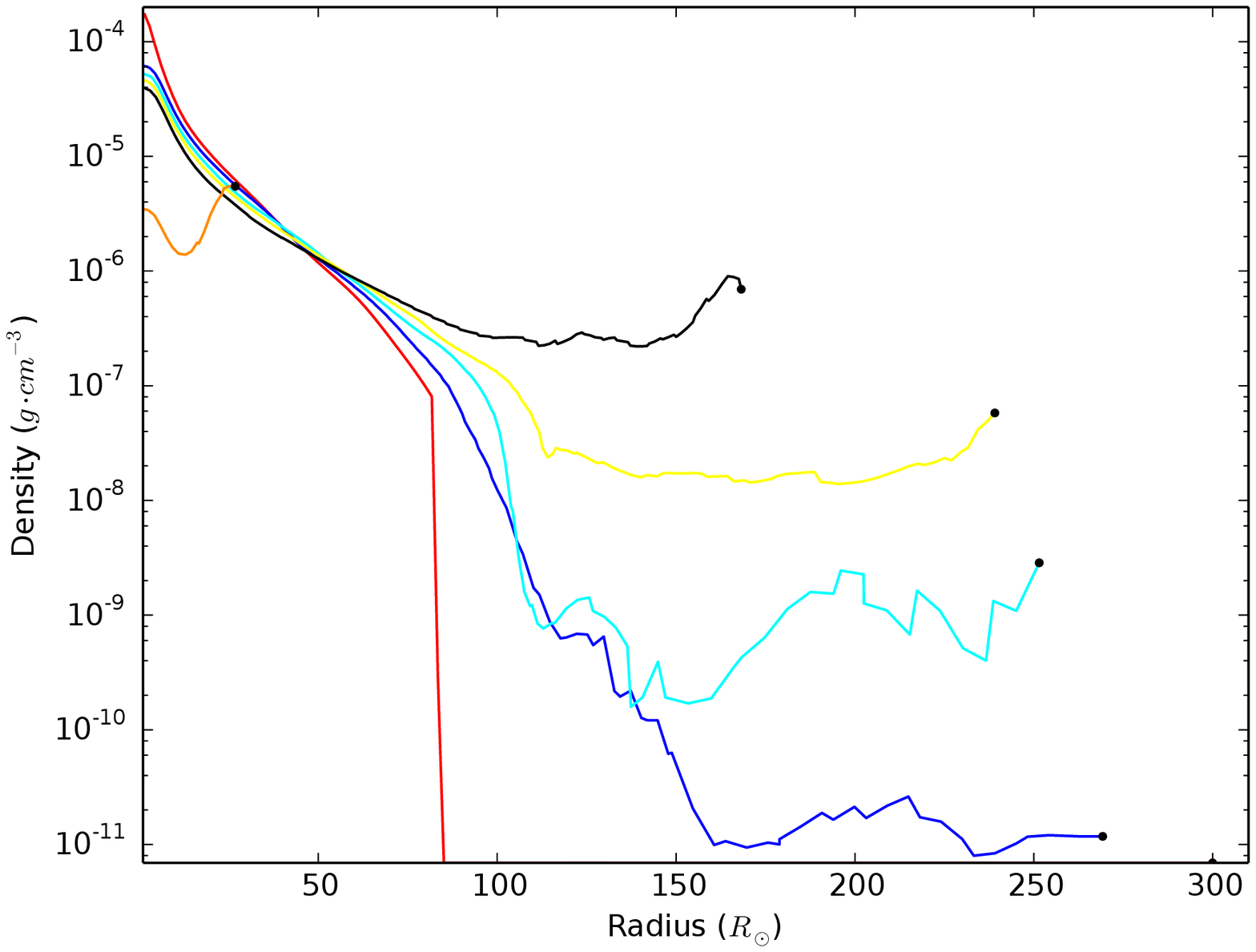}}
\subfigure[]{\label{fig:a}\includegraphics[scale=0.4]{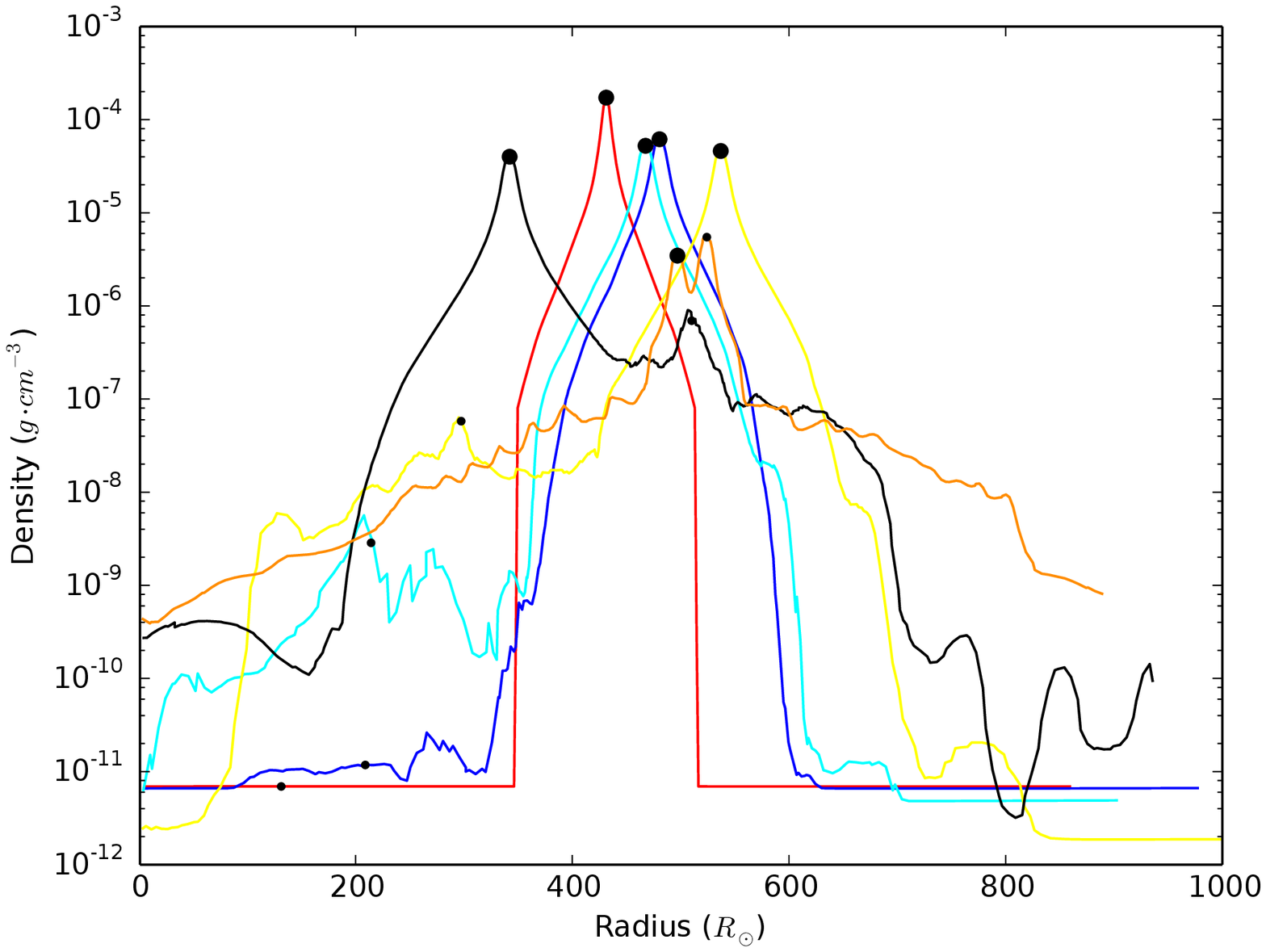}}
\caption{\protect\footnotesize{Upper panel: density profile between the the core of the primary (located at zero in the abscissa) and the companion (each black dot represents the density at the location of the companion) for the {\sc Enzo} simulation. Profiles are taken, for clarity, every $110$ dumps of the code. Colours represent times as follows: red = $0$~days, blue = $398$~days, cyan = $796$~days, yellow = $1194$~days, black = $1591$~days, orange = $1989$~days. Lower panel: same as the top panel, but extended to the whole box. The primary's core is represented by a large black dot while the companion is marked as a smaller dot.}}
\label{density_between_particles}
\end{figure}

In Figure~\ref{density_between_particles} we display the evolution of the density profile between the two cores for the {\sc Enzo} simulation, showing only the part between the particles (upper panel), or the entire computational domain (lower panel). The density profile changes smoothly at the beginning of the simulation, with the primary expanding, but it then transitions into a phase of more rapid change at the onset of the rapid in-spiral phase, when the profile flattens and then becomes U-shaped, showing peaks at the locations of both the primary's core and the companion with densities of $2.8 \times 10^{-6}$~g~cm$^{-3}$ for the primary and $4.6 \times 10^{-6}$~g~cm$^{-3}$ for the companion. The underlying density is of the order of $10^{-6}$~g~cm$^{-3}$ at $365-730$~days after the start of the simulation. These values are comparable to those of P12 (their figure 13, middle panel). 

The gas density in the proximity of the particles at the end of the simulation is high, and is unlikely to be the cause of the observed slowing down of the in-spiral. From the density profiles Figure~\ref{density_between_particles} (lower panel) it is clear that during the evolution of the system some of the envelope accumulates around the companion. The accumulation of mass is negligible until the beginning of the rapid in-spiral phase, during which it starts to increase because the companion is plunging into the denser parts of the envelope. The companion local density is a factor of a few larger than the density 10-20~\rs\ away from it. The density gradient underlying the density peak near the companion is small and likely unimportant to the in-spiral. 
 
\renewcommand*{\thesubfigure}{} 

\begin{figure}
\centering     
\subfigure[]{\label{fig:a}\includegraphics[scale=0.4]{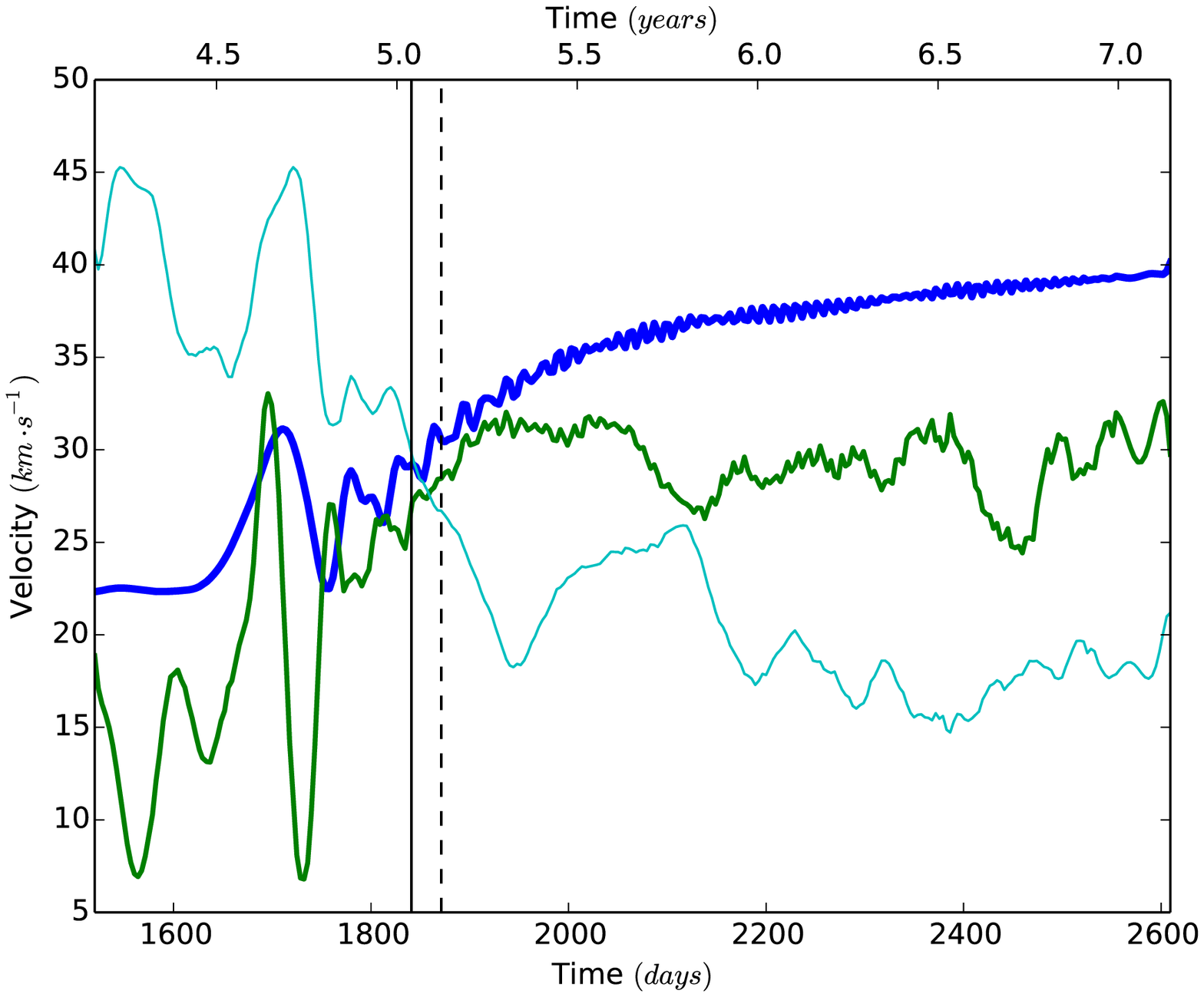}}
\subfigure[]{\label{fig:a}\includegraphics[scale=0.4]{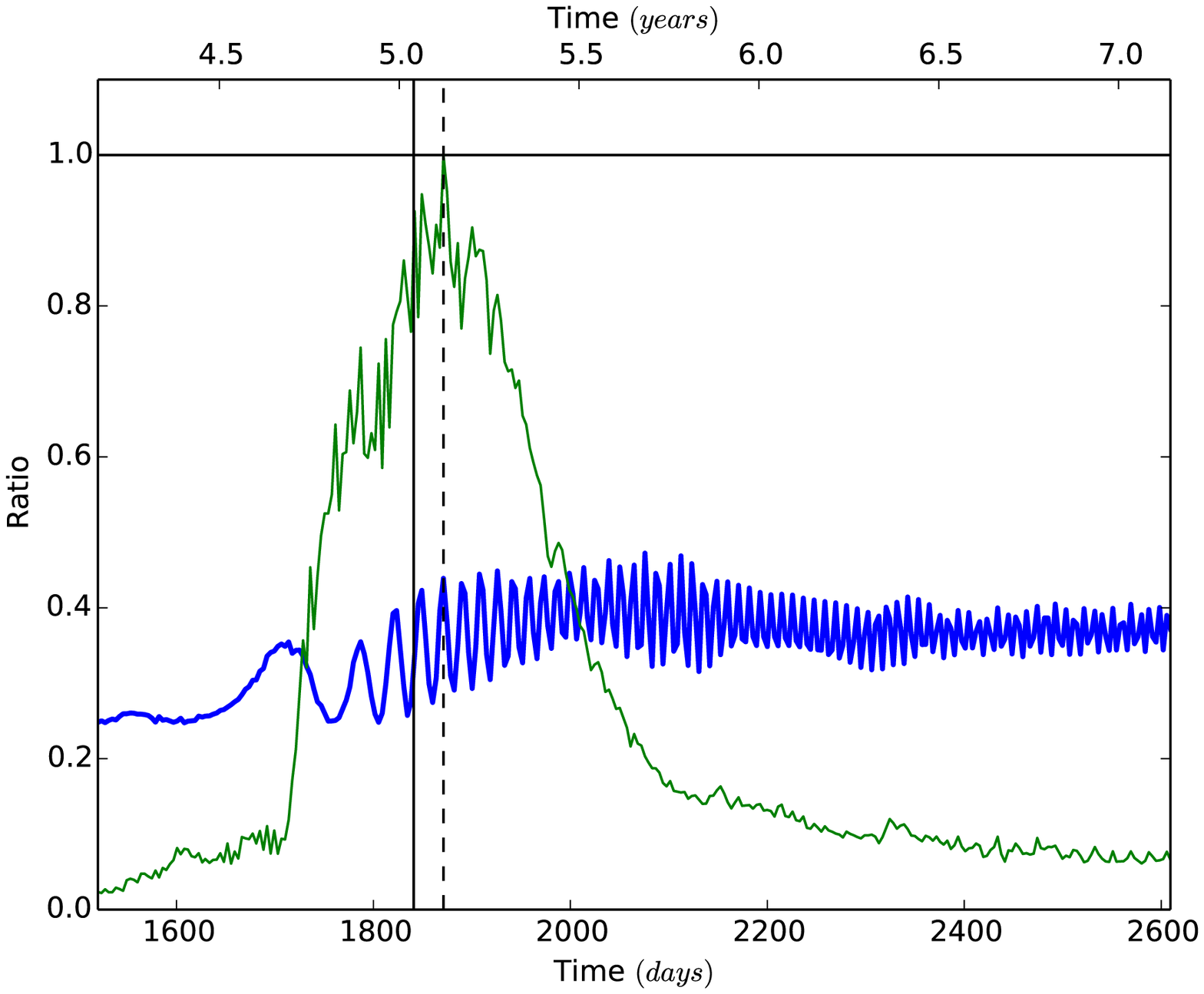}}
\caption{\protect\footnotesize{Upper panel: companion velocity (thicker blue line), local average gas velocity projected on the direction of the companion velocity ($\left< v_{\mathrm{gas,\|}} \right>$, thick green line) and local average gas velocity perpendicular to the direction of the companion velocity ($\left< v_{\mathrm{gas,\bot}} \right>$, thin cyan line) for the {\sc Enzo} simulation. The three lines are smoothed with a Savitzky-Golay filter, using $31$ coefficients and $7$th order polynomials. Lower panel: companion Mach number (thick blue line) and normalised average gas density in the companion's proximity ($\left< \rho \right>/\left< \rho \right>_{\mathrm{max}}$, where $\left< \rho \right>_{\mathrm{max}} \simeq 1.15 \times 10^{-5}$~g~cm$^{-3}$; thin green line). All plots start at the onset of the rapid in-spiral, the vertical solid lines represent the estimated end of the rapid in-spiral and the dashed ones mark the point of maximum density.}}
\label{gravodrag_efficiency_quantities_jc-300}
\end{figure}

In Figure~\ref{gravodrag_efficiency_quantities_jc-300} (upper panel) we plot the companion's speed, the average local gas velocity projected in the direction of motion of the companion and the average local gas velocity projected in the direction perpendicular to the motion of the companion for the {\sc Enzo} simulation. In Figure~\ref{gravodrag_efficiency_quantities_jc-300} (lower panel) we plot the companion's Mach number and the normalised average density near the companion. To calculate the parallel and perpendicular ambient gas velocities  we averaged the respective projections for all cells within a volume with radius 10~\rs\ from the companion. The local density was calculated by  averaging the density inside the same volume and the Mach number by averaging the gas sound speed within the same volume. 

As was the case for the simulation of P12, the entire journey of the companion is subsonic, reaching at most a Mach number of $0.47$. This is different from the simulations of \citet{Staff2016b}, where the initial part of the in-spiral was supersonic and the end of the in-spiral phase appeared to coincide with the transition between super-sonic and sub-sonic regimes. No such transition occurs here.

A regime change does however take place at the approximate time of the end of the in-spiral. This change seems to be initiated by the gas near the companion being brought into near co-rotation at approximately 1750~days. At this point, while the orbital separation is still reducing, there is still a considerable outflow (cyan line in Figure~\ref{gravodrag_efficiency_quantities_jc-300}), which eventually leads to a decrease of the local density after approximately 1870 days (dashed line in Figure~\ref{gravodrag_efficiency_quantities_jc-300}). At that point the companion's velocity's increase slows down (blue line in Figure~\ref{gravodrag_efficiency_quantities_jc-300}, upper panel) as do both components of the local gas velocity (cyan and green lines in Figure~\ref{gravodrag_efficiency_quantities_jc-300}, upper panel). By approximately 2100 days, the density has reduced so much (green line in Figure~\ref{gravodrag_efficiency_quantities_jc-300}, lower panel) that the gravitational drag is very small and the parameters of the binary change very slowly. Most of the unbinding happens at 1700~days, right after most of the orbital decay has taken place (Figure~\ref{separation_vs_time}, upper panel).

To confirm that this trend is not a result of the size of the sphere used to estimate our quantities, we carried out the same test with spheres of $5$~\rs \ and $20$~\rs. Both show results similar to Figure \ref{gravodrag_efficiency_quantities_jc-300} with the only exception that the gas velocity parallel to the companion direction of motion is overall larger and close to the companion's velocity for the smaller sphere, as expected. We also note that at the beginning of the in-spiral the local gas has a rotation velocity of 10-20~km~s$^{-1}$, which is a range of values expected for giants spun up by companions. 

\renewcommand*{\thesubfigure}{} 

\subsection{Angular momentum and energy conservation}
\label{ssec:conservation}

Energy and angular momentum were conserved by the SPH simulations of P12 at the $1$ per cent level. They did not check the conservation level of their equivalent {\sc Enzo} simulations, because of the grid nature of the code which leads to loss of mass off the simulation box and because their {\sc Enzo} simulations showed similar results to the SPH ones, which suggested a reasonable level of energy conservation.
 
As mentioned in Section \ref{sec:setup}, \citet{Staff2016a} quantified the level of energy non-conservation in grid based simulations using {\sc Enzo} and determined that conservation is improved by selecting a larger smoothing length of 3 cells rather than what was used by P12 (1.5 cells). The highest resolution in our AMR simulation is the same as the resolution in the unigrid simulations of P12. However, we have adopted the larger smoothing length of 3 cells, which must have weakened the gravitational interaction somewhat compared to the simulations of P12.

\begin{figure}
\centering     
\subfigure[]{\label{fig:a}\includegraphics[scale=0.4]{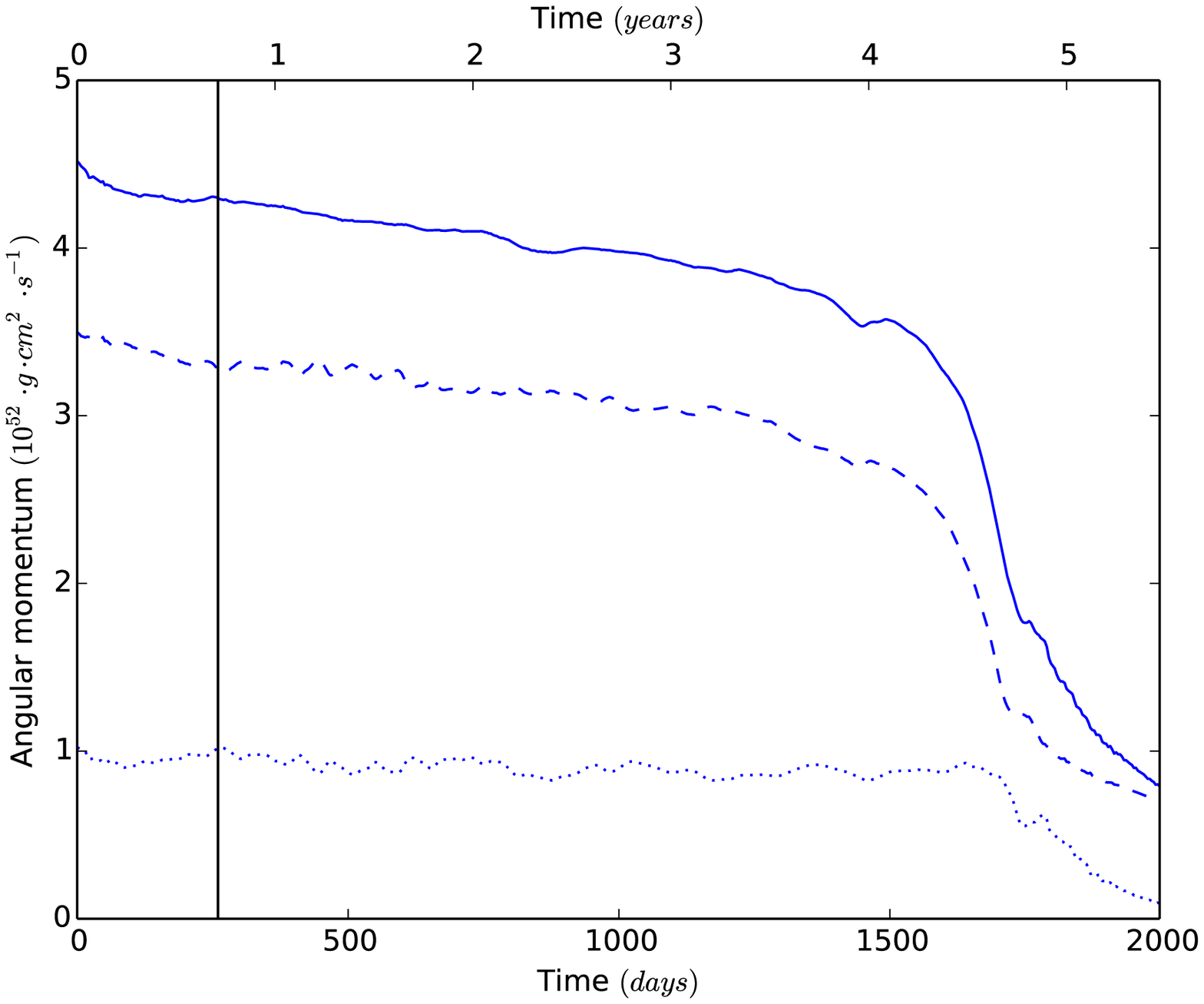}}
\subfigure[]{\label{fig:a}\includegraphics[scale=0.4]{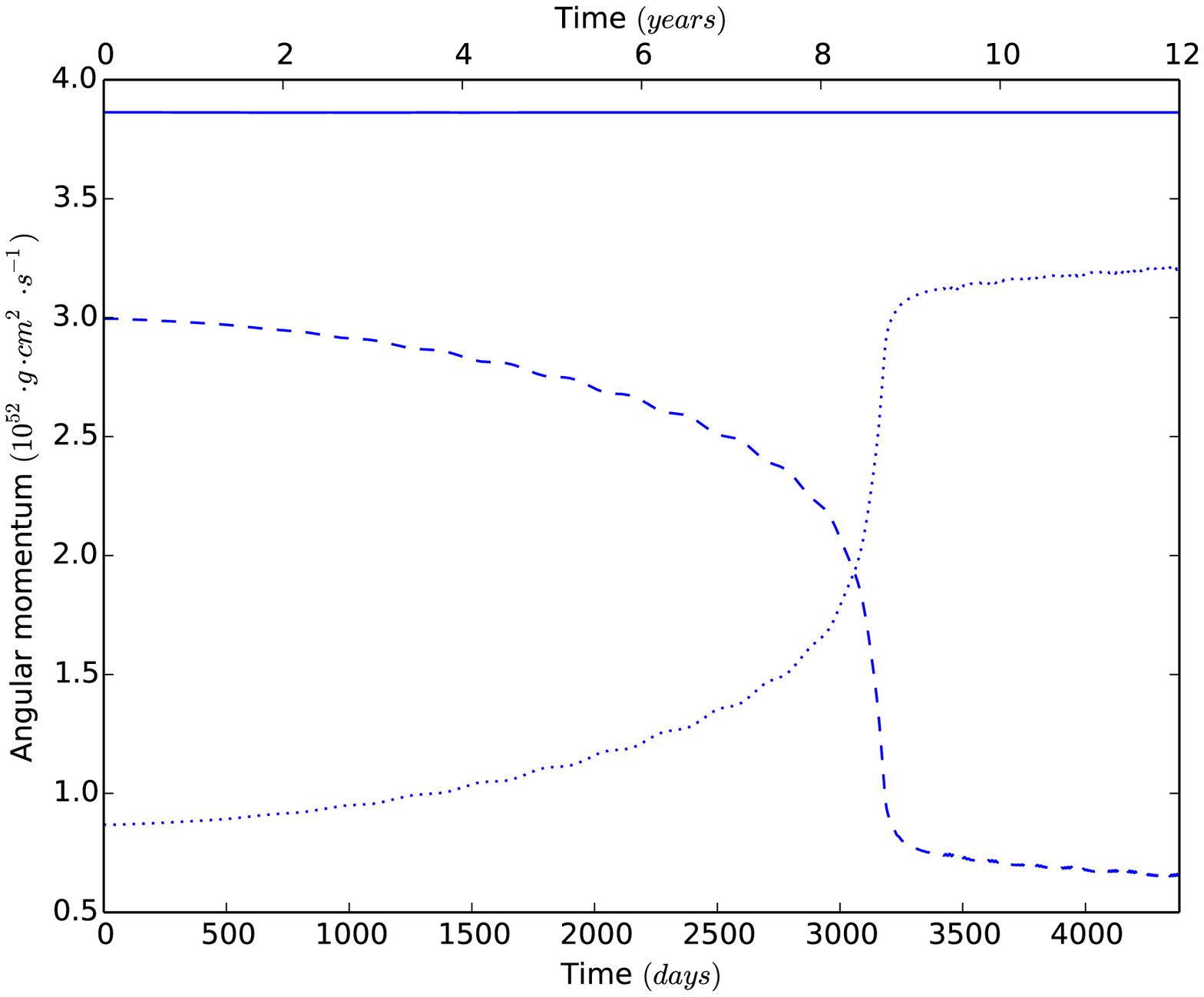}}
\caption{\protect\footnotesize{Upper panel: evolution of the $z$ component of the angular momentum with respect to the centre of mass of the system for gas (dotted line), particles (dashed line) and their sum (solid line), inside the {\sc Enzo} simulation domain. The black vertical line represents the moment when the envelope mass starts leaving the box ($\simeq 260$~days). Lower panel: evolution  of the $z$ components of the angular momentum for the {\sc phantom} simulation starting at $218$~\rs. The line styles are the same as for the upper panel.}}
\label{angular_momentum}
\end{figure}

\begin{figure}
\centering     
\subfigure[]{\label{fig:a}\includegraphics[scale=0.4]{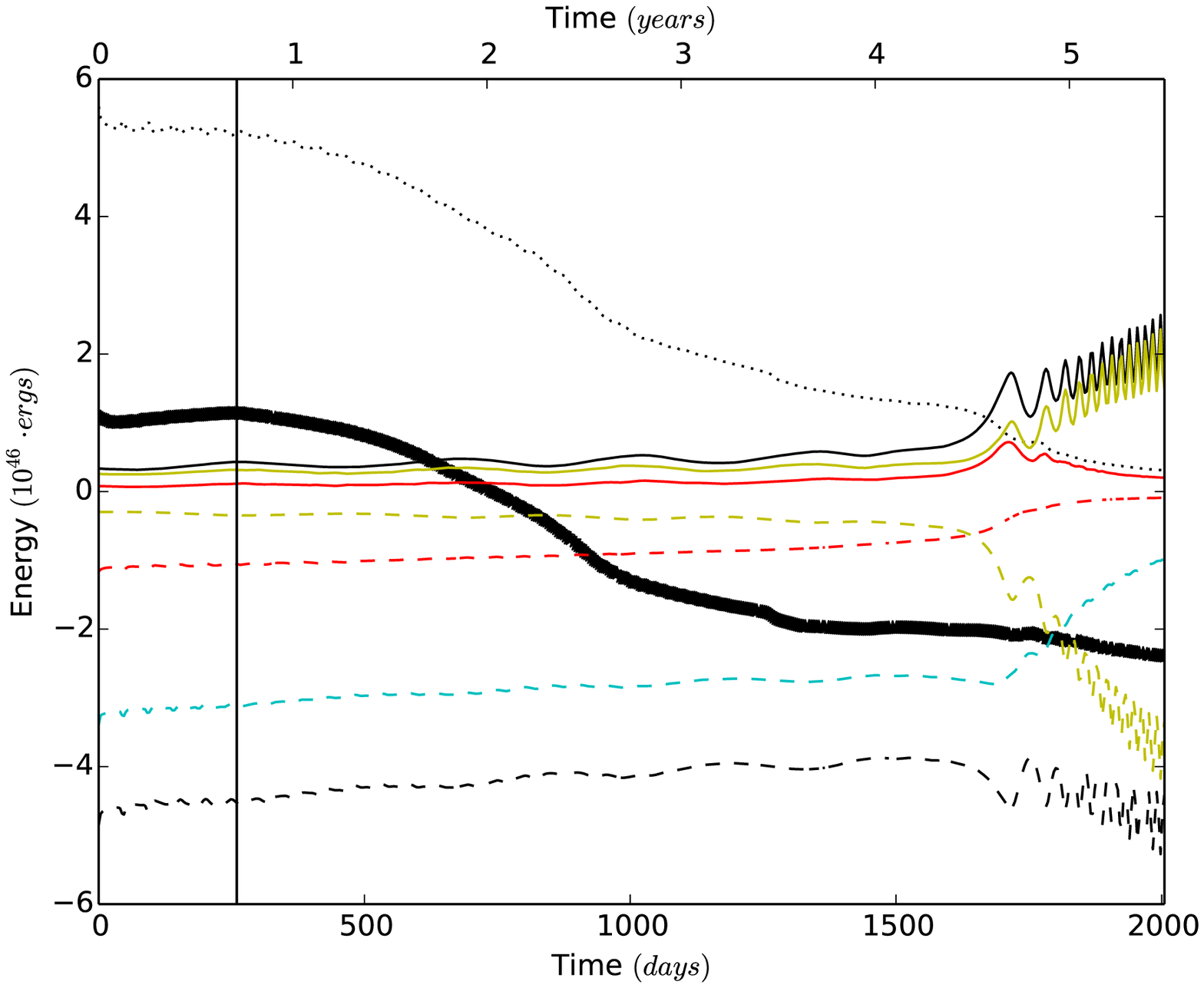}}
\subfigure[]{\label{fig:a}\includegraphics[scale=0.4]{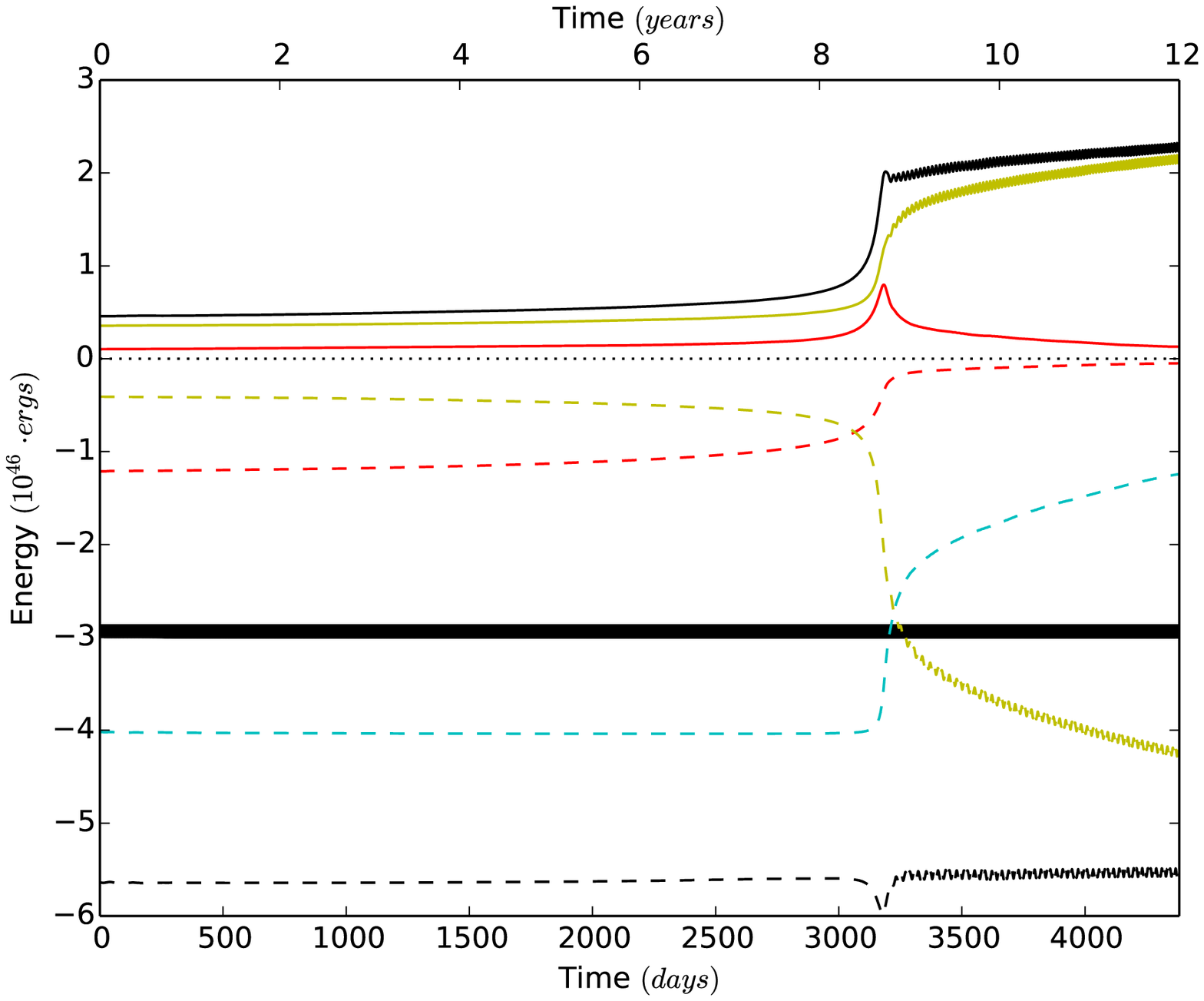}}
\caption{\protect\footnotesize{Upper panel: components of the energy as a function of simulation time in the {\sc Enzo} domain: total energy (thick black line), total kinetic energy (solid black line), total potential energy (dashed black line), total (= gas) thermal energy (dotted black line), gas kinetic energy (solid red line), gas potential energy (dashed red line), point-mass kinetic (solid yellow line), point-mass to point-mass potential (dashed yellow line) and point-mass to gas potential (dashed cyan line). The black vertical line represents the moment when the envelope mass starts leaving the box ($\simeq 260$~days). Lower panel: conservation of the components of the energy for the {\sc phantom} simulation starting at $218$~\rs. The colours are the same as for the upper panel.}}
\label{energy_vs_time}
\end{figure}

In the upper panels of Figures~\ref{angular_momentum} and \ref{energy_vs_time} we plot various components of the angular momentum and energy, respectively, in the {\sc Enzo} computational domain as a function of time. The behaviour of some of the components is driven by mass-loss out of the computational domain, which starts at $\sim$260 days (some of the low density ambient medium outflows before, but has negligible mass), but is particularly heavy during the rapid in-spiral phase.
In Figure \ref{angular_momentum} we show only the $z$ component of the angular momentum, that, as expected, dominates over the other components. Additionally, we see that most of the angular momentum resides in the point masses, with an initial value of $\sim3.5 \cdot 10^{52}$~g~cm$^2$~s$^{-1}$. Before $260$~days from the beginning of the simulation, only negligible mass and angular momentum are leaving the simulation box.  
The particles' $z$ angular momentum decreases during the in-spiral. Some of that is transferred to the gas. Five per cent of the angular momentum is lost due to non-conservation, between the beginning of the simulation and 260 days, while 10 per cent is lost over the first 3 years, a time at which substantial amount of mass starts leaving the domain. This value is larger (as expected) than for the SPH simulation of P12 and similar to the 8 per cent of \citet{Sandquist1998}, who estimated it over $\simeq 1000$~days of their simulation.  

\renewcommand*{\thesubfigure}{} 

\noindent Estimating the level of conservation of energy is even more difficult than for the angular momentum, because the low density medium filling the volume outside the star has a very high thermal energy, even if its total mass is negligible. Even before envelope mass starts flowing out of the computational domain at $260$~days, a small amount of this high energy gas flows out of the box taking with it an energy of $\simeq 1.3 \times 10^{45}$~erg (or $\simeq 11$ per cent of the initial total energy). This behaviour is clear in Figure \ref{energy_vs_time}: the total energy at the beginning of the simulation is dominated by the thermal energy of the ``vacuum" and by the potential energy between the point mass particles and the gas, with the former continuously decreasing as some of the low density medium outflows the box; this decrease is mimicked by the total energy at times greater than $260$~days. Before this threshold is passed the code conserves energy to the $4$ per cent level, similar to the result of \citet{Sandquist1998}. 

\renewcommand*{\thesubfigure}{} 

In Figures~\ref{angular_momentum} and \ref{energy_vs_time}, lower panels, we present the angular momentum and energy conservation properties for our {\sc phantom} simulation with an initial orbital separation of $218$~\rs\ (similar results were obtained for the smaller, $100$~\rs, initial orbital separation simulation). 
The angular momentum is conserved to the $0.03\%$ level over the entire simulation time, better than what obtained by P12 with {\sc snsph} (Figure~\ref{angular_momentum}, lower panel). Additionally, mass conservation in SPH simulations allows us to highlight the transfer of angular momentum from the orbit (dashed line in Figure~\ref{angular_momentum}, lower panel) to the envelope gas (dotted line in Figure~\ref{angular_momentum}, lower panel). 

Total energy (thick black line in Figure~\ref{energy_vs_time}, lower panel) is conserved in {\sc phantom} at the $0.1\%$ level, again better than  what obtained by P12 with {\sc snsph}. By comparing upper and lower panels of Figure~\ref{energy_vs_time}, one can also notice the magnitude of the contribution of the hot \virgopen vacuum\virgclose \ to the {\sc Enzo} energy budget.

Both the {\sc Enzo} (initial separation $300$~\rs) and {\sc phantom} (initial separation $218$~\rs) simulations reach the onset of the rapid in-spiral over similar, non-realistic time-scales of the order of years (Sec.~\ref{ssec:bulges}). Since {\sc phantom} conserves angular momentum well, we deduce that non-conservation in {\sc Enzo} is not the main factor driving the orbital shrinkage, though we cannot exclude that it plays a role.  

\section{Comparison with published simulations}
\label{sec:otherwork}

\begin{figure*}
\includegraphics[scale=0.45, trim=6.3cm 0.0cm 0.0cm 0.0cm]{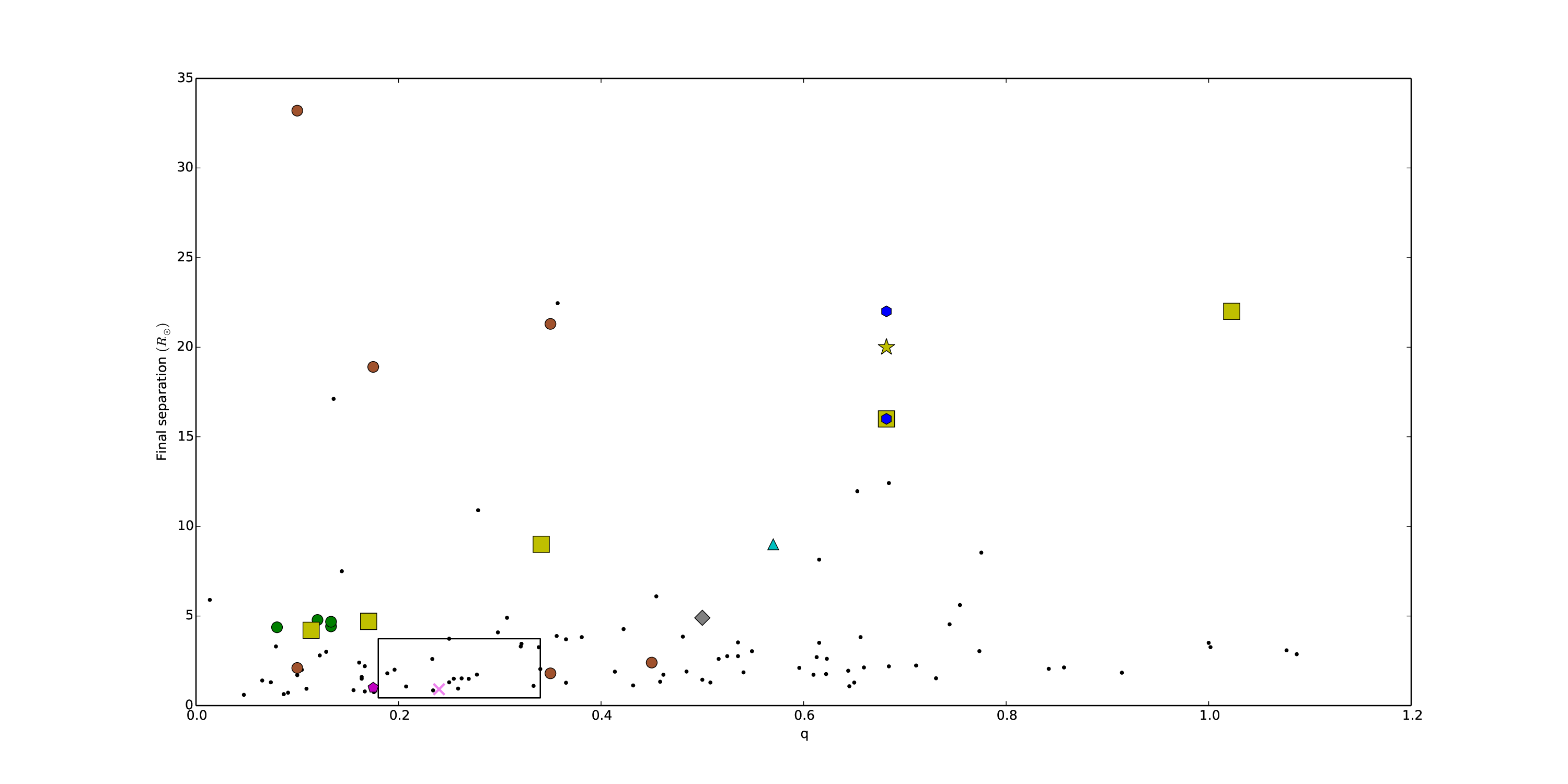}
\caption{\protect\footnotesize{Final orbital separation vs mass ratio $q=M_2/M_1$ for observed post-CE systems (\citealt{Zorotovic2010} and \citealt{DeMarco2011}, black dots) and for simulations (\citealt{Sandquist1998}, green circles; \citealt{Ricker2012}, cyan triangle - note that here we report the separation of the simulations of \citealt{Ricker2012} which is lower than reposted in \citealt{Ricker2008} where the in-spiral had not come to an end; the 256$^3$ {\sc enzo} simulations of P12 are shown as yellow squares; \citealt{Rasio1996}, magenta pentagon; the {\sc phantom} simulations carried out in this work are shown as blue hexagons; \citealt{Ohlmann2016}, grey diamond; \citealt{Nandez2015}, pink cross; \citealt{Sandquist2000}, brown circles; the {\sc enzo} simulation carried out in this work, yellow star). The rectangle encompasses all results from the low-resolution simulations of \citet{Nandez2016}.}}
\label{separation_massratio}
\end{figure*}

Here we carry out a comparison of CE simulations containing at least one giant \citep{Rasio1996,Sandquist1998,Sandquist2000,Passy2012,Ricker2012,Nandez2015,Nandez2016,Ohlmann2016}, highlighting possible trends or aspects that need further clarification. We do not include those simulations carried out by \citet{Staff2016a} that started with highly eccentric orbits. All the final results of these simulations are summarised in Table~\ref{comparison_with_previous_works_data} and we display results in Figure~\ref{separation_massratio}. All the simulations, except that of \citet{Nandez2015} are carried out with codes that include similar physics and can be more directly compared.

\subsection{Side-by-side code comparison}

The only side-by-side code comparison that can be carried out is between {\sc enzo}, {\sc snsph} and {\sc phantom} for which almost identical simulations were carried out. The comparison between the first two was already carried out by P12. Here we only add that {\sc snsph} results in final separations that are approximately 10 per cent larger than for {\sc enzo}. The relative difference does however increase for simulations with very low mass companions (0.1~\ms). 

The comparison between {\sc snsph} (simulation SPH2 in P12) and our  own {\sc phantom} simulation shows that, at the criterion point, the final separations are the same within one solar radius, while at 1000 days the {\sc phantom} separations is 10 per cent smaller, but has the same value as the {\sc enzo} simulation. We conclude that code-to-code differences for these three codes and for this parameter space are within 10 per cent for simulations with companions more massive than $\sim$0.3~\ms.

\subsection{The final orbital separation as a function of $M_2/M_1$}

Comparing the 5 {\sc enzo} simulations of P12 with each other, or their 5 {\sc snsph} simulations with each other, or, to an extent, comparing two of the simulations of \citet{Sandquist1998} for which only $M_2$ was changed, we conclude that the final separation increases for increasing value of $M_2$, for the same value of $M_1$. It is difficult to compare with the other  simulations, because although two simulations may have the same value of $q$, the binding energies of the primaries' envelope could be vastly different (but see Sec.~\ref{ssec:orb_sep_bind_en}).  Some of the simulations of \citet{Nandez2016} carried out with the same primary and different secondary masses could be used to carry out this kind of analysis were it not for the very narrow range of mass ratios available which lead to effectively the same final separation.

\citet{Sandquist1998} also compared two simulations with different primaries and the same $q$. The simulation with the more extended, lower binding envelope energy primary has a much larger final separation (see Table~\ref{comparison_with_previous_works_data}), but we did not plot it because the final separation cannot decrease much below the resolution times the particles' smoothing length and in that simulation the two values are almost they same.

The post-CE binary observations of \citet{Zorotovic2011} show that post-CE binaries with post-RGB primaries (identified by a mass smaller than 0.5~\ms) have systematically smaller separations than post-CE binaries with post-AGB primaries (which have masses larger than 0.5~\ms). They also show a marginal correlation, though statistically ``real", between secondary mass and post-CE orbital separation. The latter conclusion is in line with the simulations, though clearly the signal in the data is diluted by the range in primary masses for each secondary mass (see below).

\subsection{The final orbital separation as a function of primary mass or envelope binding energy} 
\label{ssec:orb_sep_bind_en}

The simulations of \citet{Rasio1996}, \citet{Nandez2015,Nandez2016}, \citet{Ohlmann2016} and some of the simulations of \citet{Sandquist2000} produce distinctly lower separations, at a given mass ratio, even accounting for their different values of $M_2$. We ascribe this difference to heavier and/or more compact primaries, resulting in envelopes with larger binding energies. The P12 and \citet{Sandquist1998} simulations with the most comparable values of $q$ are extremely similar, despite the fact that the lower mass for the former should promote a wider separation than the higher mass for the latter. On the other hand, \citet{Sandquist1998} simulated a more extended AGB star, which could lead to a wider separation, countering the effect of the larger primary mass. This is even more clear if we compare P12 with \citet{Sandquist2000}. In fact, for similar mass ratios, the simulations of \citet{Sandquist2000} show values for the final separations that are both smaller and larger than the giant of P12. The smaller values are obtained for primaries that are more massive and compact than the one of P12, or with the same mass but more compact than the one of P12. The larger values all result from primaries with radii at least double that of P12.

Some of the simulations of \citet{Nandez2016} were carried out with identical secondaries and a primary that evolved from the same mass star but entered a common envelope interaction at two different stages of evolution, one slightly more evolved than the other (larger core mass, larger envelope radius). From these it is clear that a doubling of the radius leads to an increase of the final separation by more than a factor of two. This is corroborated by two of the simulations of \citet{Sandquist2000}, see Table~\ref{comparison_with_previous_works_data}, carried out with identical, $1$~\ms \ primaries, but evolved to stages with smaller or larger radius ($22$ vs. $243$~\rs) which resulted in final separations that are a factor of $\sim 10$ difference ($1.8$ vs. $21$~\rs).

We do not think that the reason for the compact final configuration achieved by \citet{Nandez2015} is the extra energy source. If anything, that should have contributed to a wider separation, because of a more prompt envelope ejection. The reason is likely the more compact configuration of their RGB giant.  

\subsection{The final orbital separation as a function of giant spin at the time of Roche lobe overflow}

It could be argued that starting with a wider initial separation has, primarily, the effect of spinning up the giant, by injecting the the orbital angular momentum into the envelope. The farther the initial separation (within the limits of tidal effectiveness) the more angular momentum is available. This may in turn reduce the velocity contrast between the companion and the envelope and result in a smaller gravitational drag. However, the rotating and non-rotating simulations of \citet{Sandquist1998} reached the same final separation, indicating that the larger amount of angular momentum of their rotating star does not influence the outcome of those CE simulations. 

We conclude that the reason why \citet{Sandquist1998} did not see a difference between their rotating and non-rotating simulations is that they started the simulations with the companion on the surface of the giant. This did not give the giant time to expand before the plunge-in. In our simulations, placing the companion farther away, does not only transfer angular momentum to the giant, inducing rotation, but gives the giant time to expand. Thus, the giant gas distribution at the time of plunge-in is substantially different, being more expanded and less dense, as well as rotating. Hence, increasing the initial orbital separation leads to larger post-CE orbital separations by $25$ per cent ({\sc Enzo}) and $38$ per cent ({\sc phantom}).

\subsection{Unbound mass}

The mass unbound at the end of the simulations listed in Table~\ref{comparison_with_previous_works_data} ranges between 8 and 46 per cent (not counting the result of \citet{Nandez2015} and \citet{Nandez2016}), something that cannot be accounting for the fact that not all values were obtained with the same definition of bound mass. 

By looking at the outputs of the simulations of \citet{Sandquist1998} and P12, one could deduce that lower mass ratios ($M_2/M_1$) lead to less unbound mass. However our work, that of \citet{Rasio1996}, of \citet{Sandquist2000}, of \citet{Ricker2012} and of P12 show unbound gas masses that are overall lower than for the simulations of \citet{Sandquist1998} or the simulations of \citet{Nandez2015} that did not include recombination energy, which unbound 50 per cent of the envelope (although this is only stated in the text of that paper and no plots, nor other data are presented for that simulation). 

The simulations of \citet{Staff2016a} with a 3~\ms\ AGB star in a common envelope with 0.6-3.0~\ms\ companions have not been included in Table~\ref{comparison_with_previous_works_data} because of their high initial eccentricity, which makes them stand on their own. We note, however, that resolution tests carried out in the context of those simulations show that slightly un-converged simulations tend to unbind significantly more mass than better converged simulations. We therefore wonder whether convergence, which is seldom formally achieved in this type of time-consuming simulations, may impact the value of the unbound mass.

The impact of the recombination energy on the unbound mass was shown to be a promising avenue for further study by \citet{Nandez2015}, who derived unbound masses of almost 100 per cent, compared to 50 per cent not including recombination energy. The new simulations of \citet{Nandez2016} also unbound the envelopes using recombination energy. 
According to \citet{Nandez2016}, recombination energy should be available to the envelope in its entirety because the recombination front is at very high optical depths. This should be true also for lines, where the recombined material  becomes more optically thin, because most of the gas would be reprocessed before reaching the required temperatures. However, it is still to be clarified if including recombination energy in codes adopting the adiabatic approximation, and therefore unable to radiate, it is a valid approach in all cases. There are after all examples in nature that demonstrate that a recombination fronts developing in giant stars do not blow the star apart, for example in pulsating Miras (\citealt{Harpaz1998}).

\section{Conclusions}
\label{sec:conclusion}

In this work we have expanded on the results of \citet{Passy2012} by repeating one of their common envelope simulations, an $0.88$~\ms, RGB primary and a $0.6$~\ms\ companion, but increasing the initial orbital separation from 1 to 3 times the initial stellar radius. This is the approximate value of the orbital separation for which a tidal capture can be expected and as such it is the approximate value of the maximum angular momentum that can be delivered to the primary for such a system. We have also carried out a parallel set of simulations using the SPH code {\sc phantom} aimed at continuing code-code comparison while checking the conclusions obtained using the grid code.

We divided the evolution into a pre-contact phase, a mass transfer phase and a rapid in-spiral phase. The pre-contact phase is driven by tides. However, this phase is unrealistically short in our simulation, due to small but tidally significant oscillations of the primary star envelope set in motion by the introduction of the companion in the computational domain. 

The mass transfer and the rapid in-spiral phases are in approximate agreement with the theoretical expectations.  Starting with a larger initial separation results in a larger final separation by between 25 and 38 per cent for the set of parameters tested in this work. Based on a comparison with simulations in the literature, we conclude that this is due primarily to the stellar expansion prior to the rapid in-spiral  phase, rather than the extra angular momentum injected into the primary. 
 
In both our grid and SPH simulations, we observed that the unbinding of the mass happens in a short, bursting event which begins shortly before the rapid in-spiral phase and peaks early during it, as expected from previous work. All the unbound mass is then rapidly pushed out of the simulation box in the case of the grid simulation.
The total amount of mass unbound during the interaction is $16$ per cent of the total envelope mass, for both {\sc Enzo} and {\sc phantom}, while in the equivalent SPH simulation of P12, $10$ per cent of the envelope mass is unbound. The companion could thus eject $60$ per cent more mass than for a simulation starting with a smaller orbit, probably because by tapping the reservoir of orbital angular momentum in the wider orbit the envelope has a lower binding energy. 

Both the increase in final separation and in the amount of unbound mass discussed above are echoed by comparing our two SPH simulations started at different initial separations.  

By setting our results in the context of previous work, a new picture seems to be emerging, indicating that the discrepancy between observed post-CE separations and simulation is not as definitive as when P12 carried out their comparison, with several simulations reproducing very small final separations, even for relatively large values of the $M_2/M_1$ ratio. The strong dependence of final separation on secondary mass can only be assessed by the P12 simulations, which carried out the necessary comparison. The amount of unbound mass seems to cluster in two groups, with low ($\lesssim$15 per cent) and  high ($\gtrsim$40 per cent) values, but the reason for this difference is not clear.

\citet{Nandez2015} report to have resolved the problem of unbinding the CE by including recombination energy in their simulations, therefore their pioneering study (see also \citealt{Ivanova2015} and \citealt{Nandez2016}) constitutes a step that will have to be considered and tested further in future numerical simulations, extending the variables parameter space and simulating a wider range of astrophysical objects.

It is hoped that future simulations by different groups will attempt to clarify some of the questions above by carrying out similar simulations with a range of parameters. In this paper we have also compared the simulations with the observations previously used by P12. However, additional observations, such as those by \citet{Zorotovic2011} show new trends, which can guide parameter choices of future simulations.

\section*{Acknowledgments}
\label{sec:acknowledgments}
RI is grateful for financial support provided by the International Macquarie University Research Excellence Scholarship. JES acknowledges support from the Australian Research Council Discovery Project (DP12013337) programme and the University of Florida Theoretical Astrophysics Fellowship. OD gratefully acknowledges support from the Australian Research Council Future Fellowship grant FT120100452. JCP acknowledges funding from the Alexander von Humboldt Foundation. DP and JW acknowledge funding from ARC Discovery Project DP130102078 and DP via Future Fellowship FT130100034. This research was undertaken, in part, on the National Computational Infrastructure facility in Canberra, Australia, which is supported by the Australian Commonwealth Government, on the swinSTAR supercomputer at Swinburne University of Technology and on the machine Kraken through grant TG-AST130034, a part of the Extreme Science and Engineering Discovery Environment (XSEDE), supported by NSF grant number ACI-1053575. Computations described in this work were performed using the ENZO code (http://enzo-project.org), which is the product of a collaborative effort of scientists at many universities and national laboratories. We are also grateful for extremely helpful comments by the anonymous referee. 

\bibliographystyle{aa}
\bibliography{bibliography}{}
\bsp

\label{lastpage}

\end{document}